\RequirePackage{fix-cm}

\documentclass[twocolumn]{svjour3}

\usepackage{amsmath,amsfonts}
\usepackage{array}
\usepackage[caption=false,font=normalsize,labelfont=sf,textfont=sf]{subfig}
\usepackage{textcomp}
\usepackage{stfloats}
\usepackage{url}
\usepackage{verbatim}
\usepackage{graphicx}
\usepackage{cite}
\usepackage{xcolor}
\usepackage[most]{tcolorbox}
\usepackage{examplep}

\usepackage{hyperref}
\usepackage[table,xcdraw]{xcolor}
\usepackage{subfig}

\usepackage{comment}
\usepackage{amsmath}
\usepackage{graphicx} 
\usepackage{url}
\usepackage{enumitem} 
\usepackage{bm} 
\usepackage{upgreek} 
\usepackage{bbding} 
\usepackage{pifont} 

\usepackage{listings} 

\usepackage{tabularx} 
\usepackage{booktabs} 
\usepackage{multirow}
\usepackage{hhline} 
\usepackage{diagbox}

\usepackage{dsfont}
\usepackage[mathscr]{euscript}
\usepackage{amsfonts}
\usepackage{algorithmic}
\usepackage{graphicx}
\usepackage{textcomp}

\usepackage{pgfplotstable} 
\pgfplotsset{compat=1.16}

\usepackage{tikz} 
\usetikzlibrary{bayesnet}
\usetikzlibrary{matrix}
\usetikzlibrary{calc}
\usetikzlibrary{math}

\usepackage{adjustbox}
\def\eox{\unskip\kern 10pt{\unitlength1pt\linethickness{.4pt}$\diamondsuit${}}} 

\usepackage{easybmat} 

\usepackage{rotating}		

\usepackage[ruled,noend,linesnumbered]{algorithm2e} 

\DontPrintSemicolon     
\SetNlSty{}{}{}                
\SetAlgoInsideSkip{smallskip}   
\SetAlCapSkip{1.5mm}             
\SetInd{1.5mm}{1.5mm}             

\SetAlFnt{\small}			
\SetAlCapFnt{\small}		
\SetAlCapNameFnt{\small}

\usepackage[T1]{fontenc}
\usepackage[utf8]{inputenc}

\newcommand{\hide}[1]{} 		

\newtheorem{questionW}{Question}
\newtheorem{resultW}{Result}

{{\textit{Question \arabic{questionW}#1.}
  \stepcounter{questionW}}}
{}
\usepackage{tcolorbox}		
\tcbuselibrary{breakable,skins}		

\tcbset{examplestyle/.style={
		enhanced jigsaw,	
		colback=blue!10,	
		colframe=blue!10,	
		arc=0mm,
		boxrule=0pt,		
		left=1mm,
		right=1mm,
		left skip=-1mm,  
		right skip=-1mm, 
		top=0pt,		
		bottom=0pt,		
		breakable,		
		parbox = false,		
		before={\par\pagebreak[0]\vspace{1mm}\parindent=0pt},
		after={\par\pagebreak[0]\vspace{1mm}\parindent=0pt},
		bottomrule = 0mm,
		boxsep = 0mm,					
		topsep at break=0pt,			
		bottomsep at break=0pt,			
		pad at break=0mm,
		pad before break=0mm,
		pad after break=1mm,
		bottomrule at break=0mm,
		toprule at break=0mm,
		}}

\tcolorboxenvironment{example}{examplestyle}
\usepackage{footnote}

\tcbset{qrboxstyle/.style={
		enhanced jigsaw,	
		colback=gray!20,
		colframe=gray!40,
		arc=0mm,
		boxrule=1pt,		
		left=1pt,
		right=1pt,
		topsep at break=1mm,			
		top=1pt,		
		bottom=0mm,		
		breakable,		
		parbox = false		
		}}

\newcounter{resultboxenv}

\newsavebox{\coloredbgbox}

\usepackage{adjustbox}
\def\eox{\unskip\kern 10pt{\unitlength1pt\linethickness{.4pt}$\diamondsuit${}}}

\makeatletter
\DeclareRobustCommand*\uell{\mathpalette\@uell\relax}
\newcommand*\@uell[2]{
  \setbox0=\hbox{$#1\ell$}
  \setbox1=\hbox{\rotatebox{10}{$#1\ell$}}
  \dimen0=\wd0 \advance\dimen0 by -\wd1 \divide\dimen0 by 2
  \mathord{\lower 0.1ex \hbox{\kern\dimen0\unhbox1\kern\dimen0}}
}
\makeatother

\usepackage[normalem]{ulem} 
\newcommand\hl{\bgroup\markoverwith
  {\textcolor{yellow}{\rule[-.5ex]{2pt}{2.5ex}}}\ULon}

\usepackage{marginnote}

\renewcommand{\epsilon}{\varepsilon} 

\newcommand{\R}{{\mathbb{R}}} 

\DeclareMathOperator*{\argmax}{argmax}

\newtheorem{axiom}{Axiom}

\newcommand{\spara}[1]{\smallskip\noindent{\bf #1}}
\newcommand{\ourmethod}{\textsc{Thetis}\xspace}
\newcommand{\ituple}[1]{$\langle${\normalfont\textsf{#1}}$\rangle$} 
\newcommand{\etuple}[1]{$\langle{#1}\rangle$} 

\newcommand{\powerset}{\mathscr{P}}

\newcommand{\Dl}{\mathcal{D}} 
\newcommand{\Vals}{\mathcal{V}}  
\newcommand{\At}{\mathcal{A}} 

\newcommand{\labelf}{\lambda}
\newcommand{\gtriple}[3]{{\langle}#1,#2,#3{\rangle}}
\newcommand{\mapfun}{\Phi}
\newcommand{\score}{\sigma}
\newcommand{\semrel}{\textsc{SemRel}}
\newcommand{\Inf}{\mathrm{I}}

\newcommand{\approxtext}[1]{\ensuremath{\stackrel{\text{#1}}{\approx}}}
\newcommand{\napproxtext}[1]{\ensuremath{\stackrel{\text{#1}}{\not\approx}}}
\newcommand{\tem}{\approxtext{TE}}
\newcommand{\ntem}{\napproxtext{TE}}
\newcommand{\pem}{\approxtext{PE}}
\newcommand{\tam}{\approxtext{TR}}
\newcommand{\pam}{\approxtext{PR}}

\newcommand{\graph}{G}  
\newcommand{\nodes}{\mathcal{N}} 
\newcommand{\literals}{\mathcal{L}}  
\newcommand{\edges}{\mathcal{E}} 

\begin{document}

\title{Exploring Semantic Table Search in Data Lakes}

\author{Martin Pekár Christensen \and
        Matteo Lissandrini \and
        Katja Hose
}

\institute{Martin Pekár Christensen \at
              Aalborg University, Aalborg, Denmark \\
              \email{mpch@cs.aau.dk}           
           \and
           Matteo Lissandrini \at
              University of Verona, Verona, Italy \\
              \email{matteo.lissandrini@univr.it}
            \and
            Katja Hose \at
                Technische Universität Wien, Vienna, Austria \\
                \email{katja.hose@tuwien.ac.at}
}

\maketitle

\begin{abstract}
Discovery of relevant datasets within data lakes remains a key challenge.
Union and join search are effective techniques for discovering data when augmenting an existing table.
In exploratory search, instead, keyword search is commonly adopted to find a relatively small set of highly relevant tables matching some search terms.
In this context, semantic search is a new technique that exploits the presence of named entities in exemplar tuples used as queries and in target tables.
Hence, keyword search and entity search are able to find a diverse set of tables.
In this work, we propose \texttt{Thetis++}, which combines keyword and semantic table search
to discover relevant tables that combine the semantic signal with the presence of exact keyword matches.
Incorporating multiple paradigms and sources of information allow us to combine the best of multiple worlds.
Therefore, we study many alternative techniques for combining rankings and evaluate these for the table search task in data lakes.
We explore both unsupervised and supervised techniques, including TOPSIS.
Finally, alternative to \texttt{Thetis++} combining keyword and semantic relevance signals, we study a variant of learning-to-rank for the table search task in data lakes.
Our evaluation highlights both the limitations of many standard approaches along with some promising directions offered by learned approaches.
Furthermore, while indexes for keyword search are quite efficient, here we provide insights into how to improve the scalability of prefiltering techniques for semantic search, comparing LSH indexes with Hierarchical Navigable Small Worlds.
Our experiments show that HNSW drastically improves the runtime of semantic table search by 4.7$\times$ on average whilst retaining the high-quality tables.
Together, our findings plot the line for effective and efficient data discovery techniques for semantic data lakes.
\end{abstract}

\section{Introduction}

In recent years, data lakes have become a primary choice for storing heterogeneous data within organizations~\cite{esmailoghli2026every,schmid2019using,dibowski2021using}.
The high flexibility of data lakes allows, among others, to ingest data independently from the source, the domain, and the application for which it has been produced.
As a result, data lakes contain an \emph{increasing number of distinct, heterogeneous datasets} with varying structures, leading to a high level of complexity and diversity in both data formats and content~\cite{schmid2019using}.
This also presents unmatched challenges for organizations in terms of data integration and data exploration workflows~\cite{chapman2020dataset, Miller18}.

\begin{figure*}[ht]
    \centering
    \subfloat[Semantic data lake with tables linked to KG entities]{%
        \includegraphics[width=0.4\textwidth]{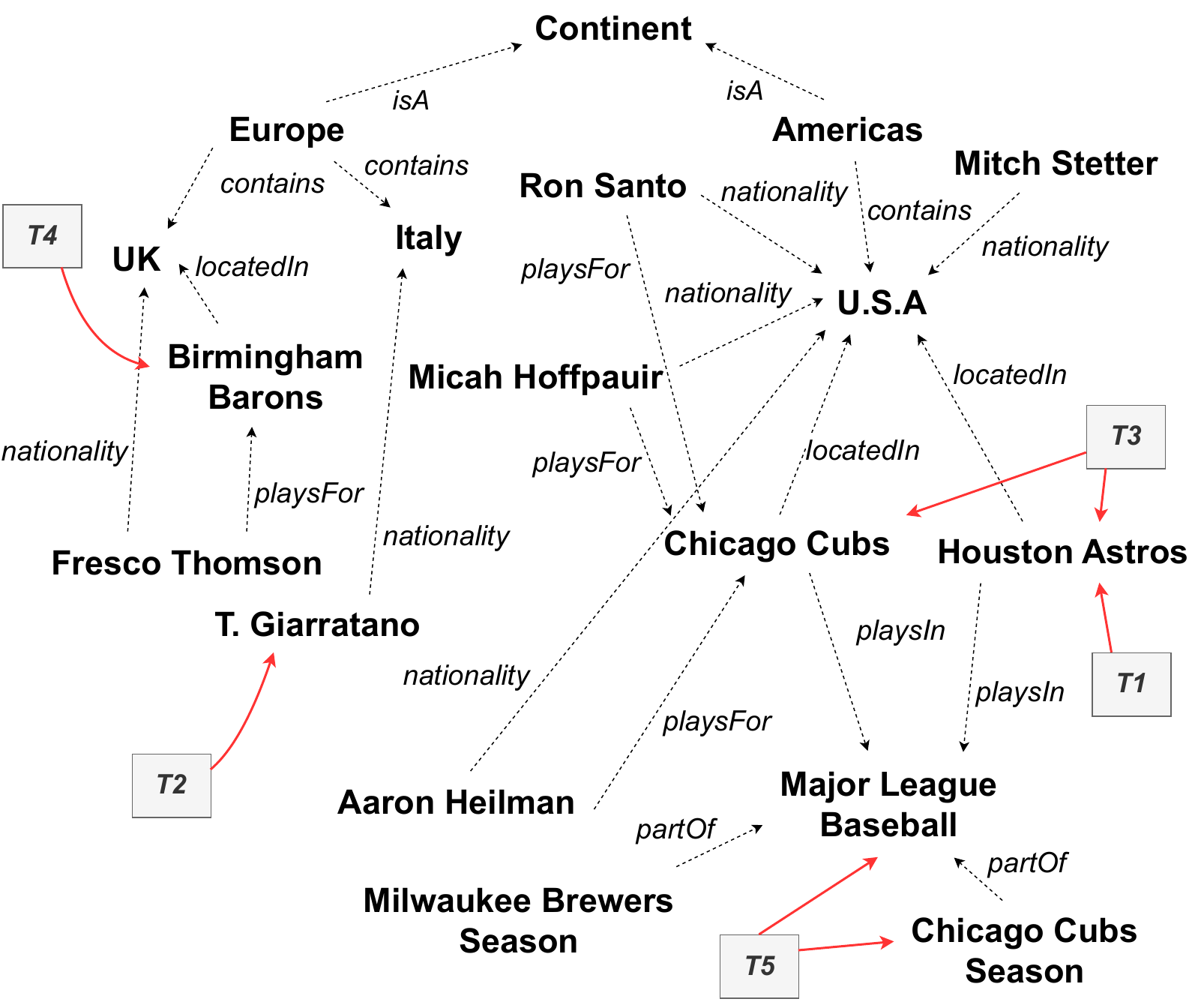}
        \label{fig:kg_tables}
    }
    \hspace{5pt}
    \subfloat[A data lake sample of baseball players, teams, and results]{%
        \includegraphics[width=0.5\textwidth]{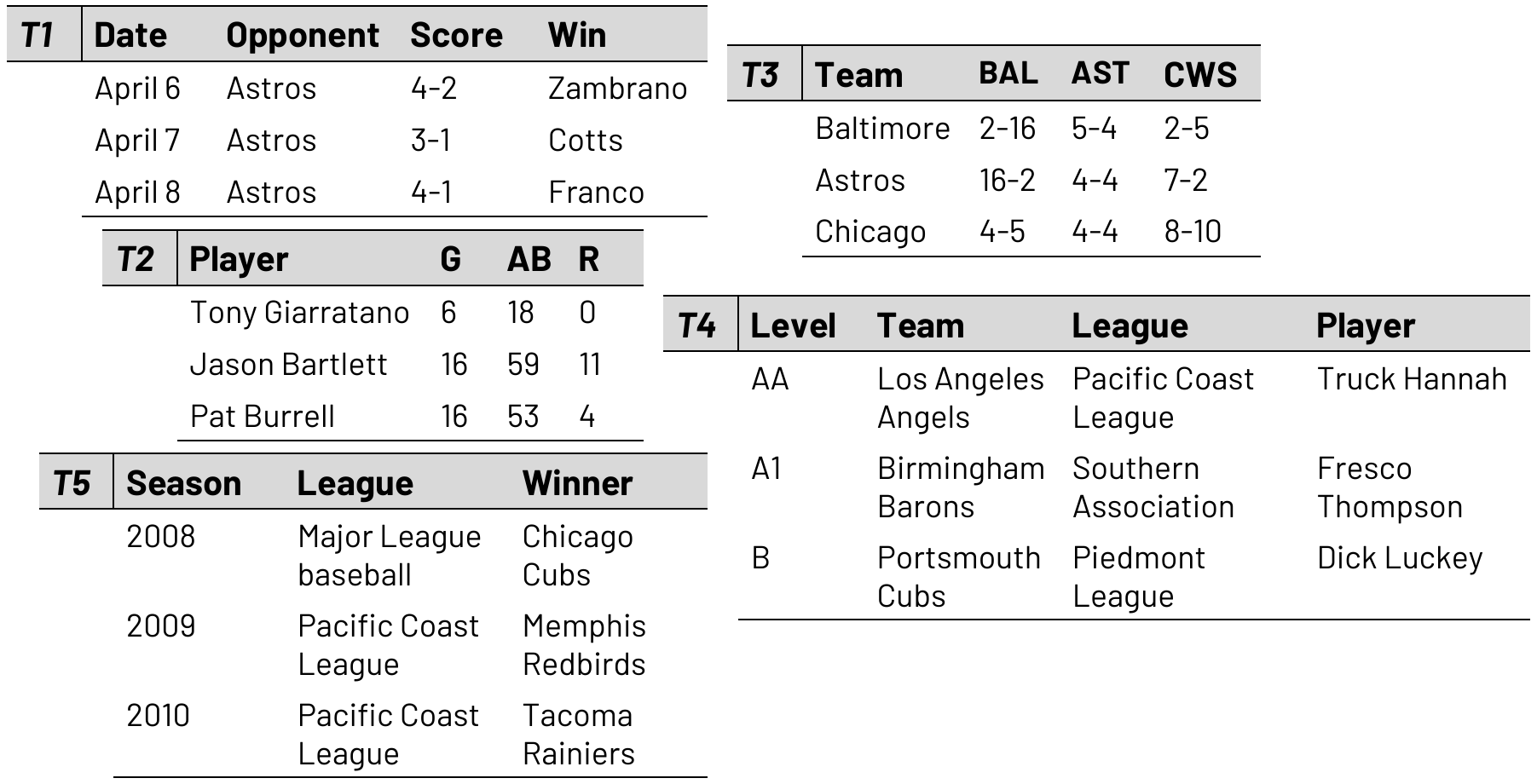}
        \label{fig:tables}
    }
    \hspace{5pt}
    \subfloat[Query to retrieve baseball players in different seasons]{%
        \includegraphics[width=0.5\textwidth]{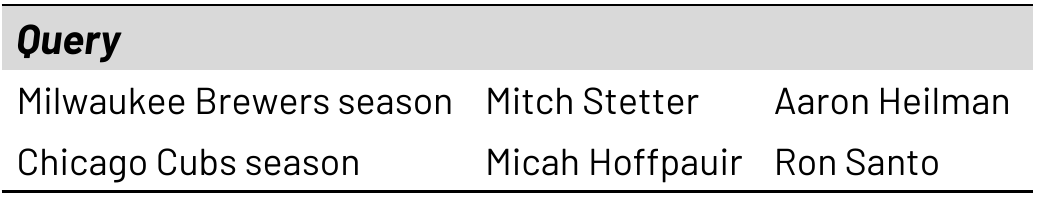}
        \label{fig:example_query}
    }
    \caption{Semantic data lake and example entity-centric search query.}
    \label{fig:sdl}
\end{figure*}

In this work, we focus specifically on the task of data discovery, where the main challenge is to find the right data to solve a given data science problem~\cite{Miller18, DataManagementSurvey}.
This is also important in RAG systems, where finding the right data enables the system to produce high-quality answers to a user query~\cite{wei2025revisitingtaskorienteddatasetsearch}.
For example, a data scientist aiming to understand a given phenomenon is frequently aware of some data describing it and still needs to search and retrieve all the other tables in the data lake with additional relevant data~\footnote{Although data lakes can store any data format, in this work, we focus specifically on CSV tables and in JSON data flattened in the form of a table.}.
However, in large-scale setups, we cannot expect data scientists to be aware of all available datasets.
Instead, data scientists may more easily be aware of entity names describing relevant items or concepts.
When a user is able to list relevant entities, they can employ an entity-centric approach, where user queries are composed of example entities of interest~\cite{mottin2016exemplar, lissandrini2018data}.
This type of data discovery requires that entities, their relationships, and concepts of interest are tracked across the entire data lake.
Knowledge graphs (KGs)~\cite{schmid2019using, sequeda2019pay, noy2019industry} are a key data model for modeling entities, relationships, and their occurrences in different datasets.
KGs form a heterogeneous network of entities, their attributes, and relationships among them - usually stored as RDF graphs~\cite{RDF}.
Hence, KGs can be employed to tackle the problem of tracking the tables connected to the entities of interest (see the example in Figure~\ref{fig:kg_tables}).
In this context, a natural question arises w.r.t. information discovery and dataset search: {\em How to find the most relevant tables for a given set of entities of interest?}

\begin{example}
Consider a data lake $\Dl$ storing a set of tables $\{T1, T2,$ $T3, T4, T5\}$ (Figure~\ref{fig:tables}) and their links to a knowledge graph as in Figure~\ref{fig:kg_tables}.
Further, assume a data scientist is looking into the performance of female basketball players across teams and is therefore looking for information describing entities represented by tuples similar to those in Figure~\ref{fig:example_query}.
Given these entity tuples as a query, our goal is to return the tables from $\Dl$ ranked by semantic relevance w.r.t. the user query.
\end{example}

\begin{figure}[t]
    \centering
    \includegraphics[width=0.8\columnwidth]{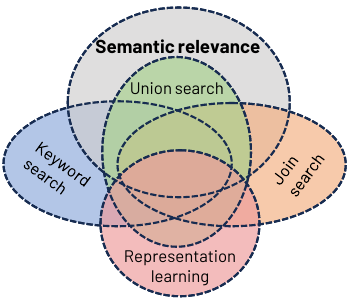}
    \caption{Relationship between table search methods' output}
    \label{fig:taxonomy}
\end{figure}

Semantic relevance search~\cite{christensen2025fantastic} is an alternative to existing table search methods and has a different goal from other search methods (see Figure~\ref{fig:taxonomy}): it consists of ranking tables by their semantic relevance to a set of example entity tuples as input query~\cite{christensen2025fantastic}.
This differs from keyword search, where tables are retrieved based on syntactic similarity to a set of input keywords; thus, retrieving a smaller set of relevant
tables (i.e., only tables $T3$ and $T5$ are retrievable in Figure~\ref{fig:tables}).
Furthermore, semantic relevance search results should be ranked based on the semantics behind the user intent instead of implicit relevance derived from the frequencies of the token matches.
To overcome these limitations, we exploit the semantic relatedness of the contents of a table through the links to the KG (as in Figure~\ref{fig:kg_tables}), enabling semantic similarity between table entities\footnote{We consider entities to be within cells; thus, each row can contain multiple entities.}.
For instance, through the links to the KG (as in Figure~\ref{fig:kg_tables}), we compare the semantic roles and similarity between entities in each table and obtain relevance signals based on the topical context of the user query as well, going beyond simple exact matches (e.g., $T6$ should be returned since it also contains female basketball players).

In the past, semantic data lakes have been limited to direct mappings between the schema of each table and the schema of a KG, i.e., making the KG a global view integrating the data of each table~\cite{ontario, squerall, constance}.
\emph{This approach requires a prohibitive amount of manual annotations to reconcile dataset schemas to match the reference KG}~\cite{sequeda2019pay, saga}.
A different approach is instead to identify entity mentions in the datasets and link only those mentions to the entities in the KG.
This entity linking process can be easily automated with little human supervision~\cite{MatchingWebTables17, tabel, lexma}.
\emph{Hence, we remove the limiting requirement of aligning all datasets to a KG, and we directly exploit entity linking annotations.}
Importantly, our approach can work even in the common case that the KGs have only partial coverage over the data lake, i.e., not all data lake entity mentions need to be linked to the reference KG. 
Finally, given entity linking annotations, we can exploit any information encoded in a KG to accommodate any semantic entity similarity.

Typical data-discovery approaches are join and union search~\cite{fan2023semanticsawaredatasetdiscoverydata, nargesian2018table}, which aim to find tables unionable or joinable with an input table query~\cite{DasSarma:2012, nargesian2019tutorial, zhu2016lshensembleinternetscaledomain}.
Nonetheless, join and union search (even when they consider semantic similarities) are not sufficient to cover all the data-discovery needs, since they still rely on some level of exact structural matches between the query and the result~\cite{fan2023semanticsawaredatasetdiscoverydata, nargesian2018table}, a requirement that semantic relevance search aims to relax.
For example, in Figure~\ref{fig:sdl}, $T6$ is not joinable or unionable with our query, so join and union search are not sufficient.
Furthermore, table search tasks, such as join and union search, require large input tables to retrieve high-quality target tables, while semantic relevance search can support queries mentioning just a handful of entities.
Further, none of these approaches leverage KGs to understand both queries and tables.
Also, representation learning-based table search approaches~\cite{tahastudy} require large input tables and are not entity-centric, and thus, their understanding of the entity semantics in tables is limited.

Hence, we propose to follow the query-by-example (QbE) paradigm, by accepting just a few example tuples as search query~\cite{Gong2023Ver, DBLP:journals/pvldb/YangHC21, Rafiei2022, zhang2020, balog2021, elexamples, bogatu2020}.
Following a query-by-example paradigm~\cite{lissandrini2018data}, the user provides an example of the data of interest, here in the form of \emph{entity tuples}, i.e., tuples containing entity mentions as in Figure~\ref{fig:example_query}, to search for relevant datasets.
Figure~\ref{fig:taxonomy} illustrates that the output of semantic relevance search provides a complementary alternative to other search approaches.

To address the need for a scalable and effective semantic search method in data lakes, we have recently introduced Thetis~\cite{christensen2025fantastic}: a semantic data lake architecture that leverages \emph{knowledge graphs} by augmenting a data lake with links between the entities in each dataset and the corresponding entities in a KG.
\emph{This allows Thetis to effectively rank semantically relevant tables, even when the KG only partially covers the data lake, making it applicable in data lakes with incomplete or inconsistent metadata.}
Thetis exploits different types of entity similarity models, both based on the set of types and on the learned embeddings of the entities.
Finally, Thetis is more easily applicable in many data lakes, as it only requires to link cell values and does not require other metadata, e.g., column headers or table captions.

Given that semantic table search and keyword search have been shown to be two alternative and effective data discovery techniques~\cite{christensen2025fantastic}, in this paper, \emph{we further extend Thetis and explore a new combined way of searching tables in data lakes}.
Currently, there are no works that aim to combine search methods into a unified approach for data discovery in data lakes.
As a result, we propose \texttt{Thetis++}: a multi-criteria ranking system that combines ranking via {\semrel}~\cite{christensen2025fantastic} and BM25 keyword search~\cite{robertson2009probabilistic} into a single, unified ranking.
\emph{We explore different ways for combining different ranking signals and study their effect on data discovery in data lakes using established information retrieval techniques}.
Among others, learning-to-rank is the employed supervised approach for similar search tasks~\cite{khamnuansin-etal-2024-mrrank, Debabrata2023}.

Moreover, in this work, we also investigate how to improve the scalability of the semantic search task.
In particular, we investigate the effectiveness of employing HNSW~\cite{hnsw} instead of LSH~\cite{zhu2016lshensembleinternetscaledomain} to pre-filter entities~\cite{christensen2025fantastic} and tables relevant to a query.

\spara{Contributions.}
Following the query-by-example paradigm~\cite{lissandrini2018data}, we present a new semantic search paradigm~\cite{christensen2025fantastic} that combines semantic and keyword search into a unified model, where the user defines an example of data of interest in the form of \emph{entity tuples} using KG entities to search for relevant datasets.
We explore this technique in comparison to other related techniques for data discovery in data lakes.

We present the following contributions:

\begin{itemize}[leftmargin=*]
    \item We formalize semantic table search in data lakes and identify four cases for semantic relevance, which we use to build 3 fundamental axioms of a semantic relevance function (Sections~\ref{sec:preliminaries} and ~\ref{sec:approach}).
    \item While existing solutions for finding tables are mostly content-based~\cite{balog2020, MannheimSearchJoin, zhu2016lshensembleinternetscaledomain, Pugnaloni2025, Lee2026} or metadata-based~\cite{brickley2019, fernandez2019dataset, constance, balog2021} we identify a clear gap for semantic search methods and methods that combine multiple relevance signals in data discovery (Section~\ref{sec:related_work}).
    \item We propose \texttt{Thetis++} that combines semantic and keyword search into a unified table search system following the query-by-example paradigm to study their various multi-criteria ranking techniques and their opportunities (Section~\ref{sec:approach} and~\ref{sec:exact}).
    \item We design a search framework for unifying multiple search methods into a single model that combines exact matches from keyword search and the semantically relevant tables (Section~\ref{sec:combined_search}).
    \item We present a search space pre-filtering technique that exploits nearest neighbor vector search to prefilter irrelevant tables (Section~\ref{sec:approximate}).
    \item We evaluate \texttt{Thetis++} against a novel, real-world benchmark~\cite{stsd} consisting of a corpus of 238,038 and 457,714 Wikipedia tables~\cite{tabel, bleifuss2021structured} from 2015 and 2019, respectively, and provide in-depth insights for table search in data lakes, identifying promising new directions to combine different relevance signals (Section~\ref{sec:evalution}).
\end{itemize}

In the following,
Section~\ref{sec:preliminaries} presents the preliminaries required to understand combined keyword and semantic table search in a semantic data lake.
Section~\ref{sec:related_work} presents the related work.
Section~\ref{sec:approach} provides the theory behind semantic table search, including the axioms of our semantic relevance score, and Section~\ref{sec:exact} presents the semantic table search algorithm.
Section~\ref{sec:combined_search} describes our algorithm for combining keyword and semantic search, followed by Section~\ref{sec:approximate} that presents our search space prefiltering technique.
Finally, we perform an evaluation in Section~\ref{sec:evalution} and conclude in Section~\ref{sec:conclusion}.
\section{Preliminaries} \label{sec:preliminaries}

\begin{figure*}[ht]
    \centering
    \includegraphics[width=\textwidth]{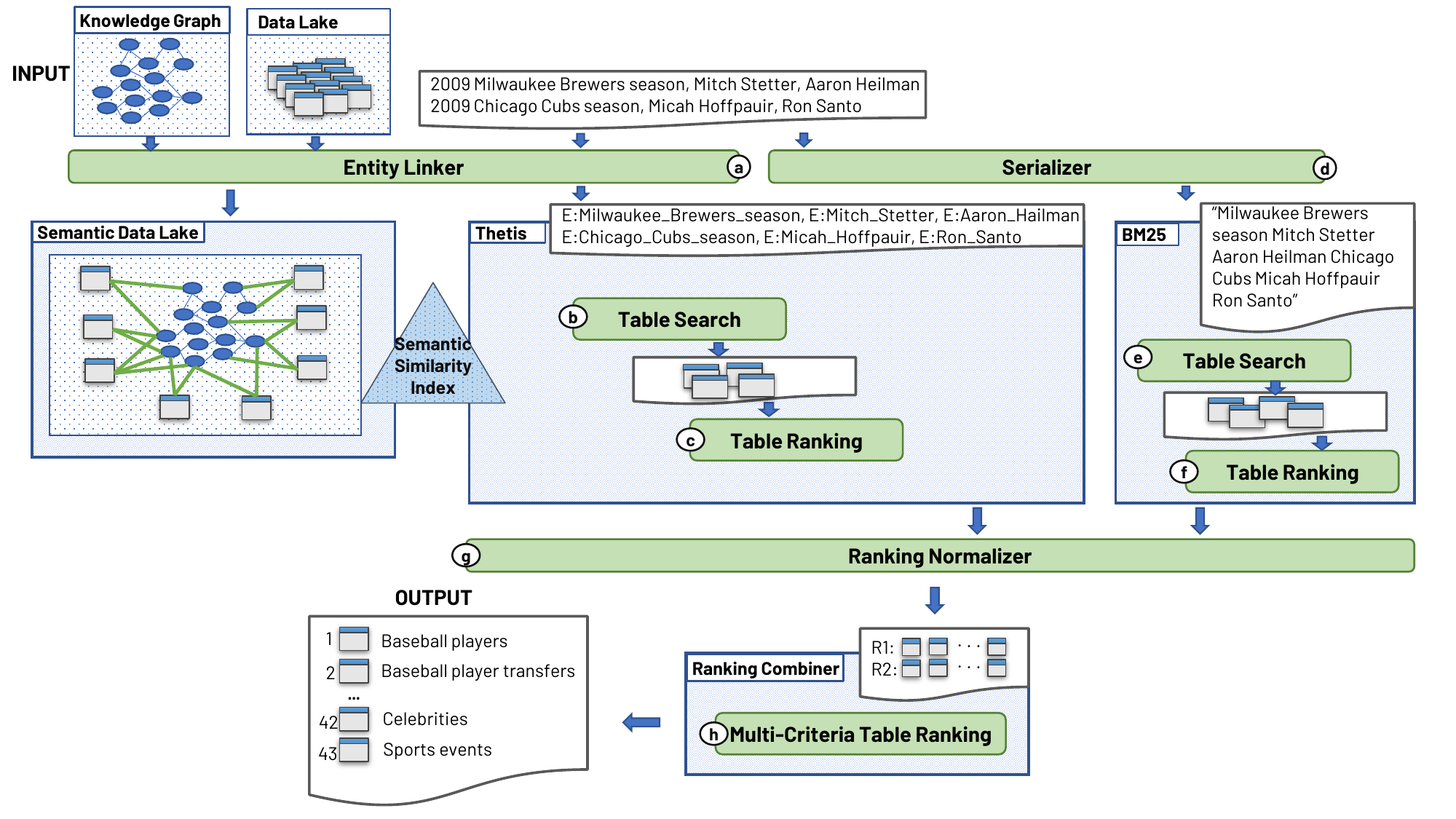}
    \caption{Architecture of \texttt{Thetis++}}
    \label{fig:overview}
\end{figure*}

Here, we present a few fundamental concepts: data lakes, knowledge graphs, semantic data lakes, semantic table search, and multi-criteria ranking~\cite{christensen2025fantastic}.
Then, we present our previous problem of semantic table search and extend this with the problem of multi-criteria ranking for combining semantic and keyword search approaches for table search in data lakes.

\subsection{Data Lake}
We define a data lake $\Dl$ as a set datasets in the form of tables: ${\Dl}{=}\{{T_1}, {T_2}, {...}, {T_n}\}$ (see Figure~\ref{fig:sdl}b).
Each table $T_{i}{\in}\Dl$ contains a set of tuples ${T_i}{=}\{{t_1}, {t_2}, {...}, {t_k}\}$ (presented as rows) with the same schema, i.e., they contain the same set of attributes $\At_i$.
That is, given a set of attribute names $\At$, each tuple $t_{j}{\in}T_i$ is described by the same list of attributes $Attr(T_i){=}\langle{a_1, a_2, {\dots}, a_m}\rangle$, with $Attr(T_i){\subseteq}{\At}$ (e.g., $Attr($\textsf{T1}$){=}\langle$\textsf{Date}, \textsf{Opponent}, ${\dots}\rangle$).
Hence, by abuse of notation, we write ${\forall}{T_i}{\in}{\Dl}$, ${\forall}{t_j, t_l}{\in}{T_i}$, $Attr(t_j){\equiv}Attr(t_l){\equiv}{Attr(T_i)}$, to say that any two tuples in $T_i$ have the same set of attributes.

Each attribute of a tuple $t_j{\in}{T_i}$ is assigned a value
from an infinite countable set of values $\Vals$ (containing both numbers and strings, plus a special value $\bot$ for the null value).
For instance, considering $t_1$ shown as the first row in table $T_2$ (Figure~\ref{fig:sdl}b), the value of the attribute \textsf{Player} for $t_1$ is ``\textsf{Tony Giarratano}''.
Finally, for simplicity, a specific attribute value in a tuple is also called a cell and identified by the attribute (the column) and the tuple identifier (the row).

Note that no information is given about referential constraints in data lakes, and therefore, there is no explicit semantic connection between table values~\cite{nargesian2018table}.

\subsection{Knowledge Graphs}
To integrate heterogeneous information, especially in an enterprise environment~\cite{schmid2019using, sequeda2019pay}, it is common to adopt a unified semantic model~\cite{Limaye:2010, Poggi:OBDA, sequeda2019pay}.
This translates into building a Knowledge Graph (KG) where entities, concepts, and attributes, along with their relationships, are described within a unified, machine-readable, information network~\cite{noy2019industry}.
Thus, a Knowledge Graph (KG) is a labeled directed graph ${\graph} = {\gtriple{\nodes}{\edges}{\labelf}}$~\cite{RDF, noy2019industry, colazzo2014rdf}, where entities, concepts, and their attributes are described as nodes $\nodes$, and the relationships among them are described as edges $\edges$.
Moreover, nodes and edges are annotated with a set of labels, i.e., human-readable literals, $\literals$ by a mapping function $\labelf: \nodes\cup\edges\mapsto\literals$.

A KG describes in a unified model both the entities of interest (e.g., people) along with relevant attributes (e.g., age), as well as their categorizations (e.g., types of profession, also called a taxonomy).
This allows the modeling of complex connections between many different types, entities, and concepts in a KG, e.g., a multi-step path connecting an athlete in a specific sport with athletes in other sport disciplines.

\subsection{Semantic Data Lake}
Numerous works match data lake tables to the structure and content of reference KGs~\cite{dibowski2021using, schmid2019using, bagozi2019, ontario}.
However, they require a reliable mapping between KG schema and the schema of each data lake table.
Current solutions performing this mapping automatically are still limited~\cite{balog2020}.
Therefore, obtaining a semantic data lake via this type of mapping requires significant manual effort~\cite{sequeda2019pay}.
Entity linking, on the other hand, aims to match entity mentions in tables to KG entities, which is a much easier task, and there exist several automatic solutions~\cite{tabel, MatchingWebTables17, Limaye:2010, Yakout:2012}.
We therefore propose to exploit entity linking between table values and KG entities, e.g., we map \textsf{Chicago} in $T1$ to the entity \textsf{Chicago Cubs} in a KG.

The resulting integration of a data lake with a KG through only entity linking constitutes the definition of a \emph{Semantic Data Lake}~\cite{christensen2025fantastic}:

\begin{definition}[Semantic Data Lake]\label{def:semDL}
A Semantic Data Lake is defined by
(1) a data lake repository ${\Dl}{:}\{{T_1}, {T_2}, {...}, {T_n}\}$,
(2) a knowledge graph ${\graph}{:}{\gtriple{\nodes}{\edges}{\labelf}}$, and
(3) a partial mapping function from data lake values to KG entities
${\mapfun}{:}\Dl{\times}\At{\times}{\Vals}{\mapsto}{\nodes}$
with its inverse from entities to the power set of data lake values ${\mapfun}^{-1}{:}{\nodes}{\mapsto}{\powerset}(\Dl{\times}\At{\times}{\Vals})$.
\end{definition}

Specifically, the two mapping functions map only some of the values and the entities in the KG and in the data lake tables.
Therefore, our system can adapt to the case when the KG does not describe all the information in the data lake (e.g., in the case when the KG does not contain some entities appearing as values in the data lake).
This is particularly important, as datasets are usually added in a data lake in a highly dynamic fashion, and thus, the data ingestion capacity of our system is not hindered by requiring exact and complete mapping.

\subsection{Problem Formulation}
We identify a new task enabled by table search in data lakes~\cite{christensen2025fantastic}, defined via our definition of a Semantic Data Lake, and by keyword search, combining these methods into a unified approach for data discovery.
Given a set of entity tuples, we aim to retrieve from the semantic data lake a ranked list of tables relevant for the given input.
We assume as input a query which is a set of entity tuples $Q$:$\{{t_1}, {t_2}, {...,} {t_k}\}$ where each entity tuple $t_i$ is a list of entities from the KG ${\graph}$ of the form $\langle e_1, e_2, ..., e_n\rangle$, with $e_i{\in}{\nodes}$.
Note that, in the simplest case, $Q$ can also be composed also by tuples containing a single entity, i.e., $Q:\{{t_1}{:}{\langle}{e_1}{\rangle}, {t_2}{:}{\langle}{e_2}{\rangle}, ..., {t_k}{:}{\langle}{e_k}{\rangle}\}$.
Then, we require the definition of a \emph{semantic relevance score} ($\semrel_{\graph}({Q},{T}){\in}{[0,1]}$), which, given a user query $Q$, assigns a relevance score to each table $T$ in the data lake depending on the relatedness, computed within $\graph$, between the entities in the query and the entities in the table.
For example, $\semrel$ should compute a higher relevance score to tables describing basketball teams than to tables describing cities when comparing them to the query entity \textsf{Seattle Storm} (Figure~\ref{fig:tables}).

Note that, if the relevance score ${\semrel}(Q,T)$ for a query $Q$ and a table $T$ is zero, then the table is irrelevant and should not be returned.
However, since it is possible that the number of tables assigned a non-zero relevance score for the query is too large, we return (without loss of generality) the top-$K$ tables according to ${\semrel}$.
Therefore, we define an appropriate {\semrel} relevance score and devise efficient algorithms to retrieve tables with the highest relevance score, without computing relevance scores for all data lake tables at runtime.

\begin{problem}[Semantic Table Search] \label{prob:semantic_table_search}
    Given a semantic data lake $\langle {\Dl},{\graph},{\mapfun} \rangle$, and a set of entity tuples $Q$ as input, extract from ${\Dl}$ a ranked list $R_S$ of relevant tables describing entities semantically related to $Q$ according to their semantic relevance score $\semrel_{\graph}(Q,T)$,  s.t., ${\forall}T{\in}{\mathcal{T}'}$: ${\semrel}_{\graph}(Q,T){>}0$.
\end{problem}

Multi-criteria ranking is employed to rank a set of items by aggregating multiple criteria (criteria here also refers to relevance scores) instead of just a single one~\cite{zheng2022, Ceballos2016, puvska2024introducing}.
Given rankings $R_1, R_2, ..., R_n$, we define the overall aggregated ranking as $R_0$ derived from the aggregation function $R_0 = f(R_1, R_2, ..., R_n)$~\cite{zheng2022}.
Generally, we can express rankings $R_1^k, R_2^k, ..., R_n^k$ as a matrix $R^{K{\times}n}$, where $k$ is the top-$K$ retrieved number of tables and $n$ is the number of search engines (in this case, $n = 2$ since we combine Thetis as representative semantic search~\cite{christensen2025fantastic} and BM25 as concrete implementation for keyword search~\cite{robertson2009probabilistic}).
Hence, we define the multi-criteria ranking combiner function as $f: R^{K{\times}n}{\mapsto}R_0$.
The relevance values belonging to $R_i$ might be of a different scale than those in $R_j$.
Therefore, due to the possible incompatibility in scales, relevance score normalization is essential, such that each relevance value from $R_i$ and $R_j$ is compatible, i.e., ${\forall}x_i{\in}R_i, x_j{\in}R_j: x_i, x_j{\in}[0, 1]$.

Since semantic relevance is a relevance score complementary to other relevance scores, we require a definition for combining search methods rankings given a user query into a single, unified ranking.
An overview of this process is illustrated in Figure~\ref{fig:overview}.
As a result, we define the problem of combining keyword and semantic table search as follows.

\begin{problem}[Multi-Criteria Table Search] \label{prob:combined_table_search}
    Extract from ${\Dl}$ a ranked list $R_K$ of relevant tables of entities syntactically related to $Q$ by keyword search and a ranked list $R_S$ of semantically relevant tables to $Q$ by semantic search, and combine $R_S$ and $R_K$ into a single, unified ranking.
\end{problem}

\section{Related Work} \label{sec:related_work}

\begin{table*}[!tb]
\centering
\definecolor{myBlue}{RGB}{55, 126, 184}
\definecolor{myGreen}{RGB}{77, 175, 74}

\newcommand{\cding}[2]{\textcolor{#1}{\ding{#2}}}
\newcommand{\yes}{\cding{myGreen}{52}}
\newcommand{\no}{\cding{red}{55}}
\newcommand{\req}{\textcolor{red}{\textsf{\textbf{!}}}}

\setlength{\belowrulesep}{0pt}
\setlength{\aboverulesep}{0pt}

\setlength{\abovecaptionskip}{2mm}
\caption{Dataset search systems features~\cite{christensen2025fantastic}: fulfilled (\yes), partial (*), and missing (\no) properties to support semantic table search}
\label{tbl:relatedwork}
\begin{adjustbox}{max width=\linewidth}
\begin{tabular}{p{4.3cm}|c|ccc|ccc}
                &                      \multicolumn{1}{c}{\textbf{}}            & \multicolumn{3}{c}{\textbf{Output Tables}}                                     & \multicolumn{3}{c}{\textbf{Searching technique}} \\
\textbf{Method} & \textbf{Query Type}   & \textbf{Augmenting} & \textbf{Matching} &  \textbf{Related}  & \textbf{Metadata} & \textbf{Content} & \textbf{Semantic}   \\
\midrule
Aurum~\cite{fernandez2018aurum}         & Keyword       & \yes & \yes & \yes          & \yes & \yes & (*) \\
Auctus~\cite{auctus}                & Keyword       & \yes & \yes & \no           & \yes & \no & \no  \\
BM25~\cite{robertson2009probabilistic}  & Keyword       & \no  & \yes & \no           & (*)  & \yes & \no \\
ECAT~\cite{da2025} & Keyword & \yes & \yes & \yes & \no & \yes & \yes \\
Google Dataset Search~\cite{brickley2019, google_dataset_search} & Keyword      & \no & \yes & \no           & \yes & \no & \no  \\
OCTOPUS~\cite{cafarella2009data}           & Keyword       & \no & \yes & \no         & \yes & \yes & \no \\
\midrule

ARDA~\cite{chepurko2020ardaautomaticrelationaldata}              &  Generic table    & \yes & \no & \no          & \no& \yes & \no \\
BareTQL~\cite{Rafiei2022}                      &  Generic table    & \yes & \yes & \yes         & \yes &\yes & \no \\
D$^3$L~\cite{bogatu2020}               &  Generic table    & \yes & \yes &  (*)         & \yes &\yes & (*) \\
DICE~\cite{rezig2021dice, fernandez2018aurum}                      &  Generic table    & \yes & \yes & \yes         & \yes &\yes & \no \\
JOSIE~\cite{josie}                   &  Generic table     & \yes & \yes & \no          & \no & \yes & \no \\
JUNEAU~\cite{zhang2020}             &  Generic table    & \yes & \no & (*)      & \yes &\yes & \no  \\
MATE~\cite{esmailoghli2022matemultiattributetableextraction}              &  Generic table     & \yes & \no & \yes          & \no & \yes & \no \\
Proximity~\cite{alserafi2020keeping}   &  Generic table    & \no & \yes & \no         & \yes &\yes & \no   \\
QCR~\cite{santos2022}                   &  Generic table     & \yes & \yes & \no          & \no & \yes & \no \\
SANTOS~\cite{santos}                   &  Generic table      & \yes & \yes & (*)          & \no & \yes & (*) \\
SEMPROP~\cite{fernandez2018}        & Generic table       & \yes & \yes & \yes         & \yes & \yes & (*) \\
Starmie~\cite{fan2023semanticsawaredatasetdiscoverydata}             & Generic table        & \yes    & \yes  & (*)     & \no    & \yes & (*) \\
TableRank~\cite{liu2007tablerank} & Generic table & \no & \yes & \no & \yes & \no & \no \\
Table Union~\cite{nargesian2018table}  &  Generic table    & \yes & \yes & \no         & \no & \yes & \no \\

\midrule
DS4DM/RapidMiner~\cite{lehmberg2017winte, lehmberg2017stitching, rapidminer}                                       & Entity table      & \yes & \yes & \no          & \yes &\yes & \no \\
S3D~\cite{Galhotra2020}     &  Entity table     & \yes & \yes & \no        & \yes & \yes & (*)  \\
TURL~\cite{turl}                  & Generic table     & \yes & \no  & \no        & \no & \yes & \no \\
\midrule
Thetis~\cite{christensen2025fantastic}                              & Entity tuples     & \yes & \yes & \yes        & \no &\yes & \yes \\
\bottomrule
\end{tabular}
\end{adjustbox}
\vspace*{-8pt}
\end{table*}

In exploratory cases, \emph{entity-centric queries are quite common as they allow effective example-based exploratory search~\cite{lissandrini2018data}}, and we therefore focus on example-based approaches, in which the user employs examples of information they wish to retrieve.
We identify three categories of methods: (a) \emph{matching}, where the content of target datasets match the query content; (b) \emph{augmenting}, where the target datasets add additional tuples to the query with the same semantics or add attributes to the query tuples; and (c) \emph{related}, where the target datasets add additional contextual information to the query tuples~\cite{christensen2025fantastic}.
However, \emph{semantic data lakes are still unexplored} in the area of data discovery, and scalable semantic-aware search has not yet received sufficient attention~\cite{christensen2025fantastic}.

\subsection{Table Search in Data Lakes} \label{sec:soa_ts}
When searching a data lake, there is a growing need for search capabilities allowing scientists and analysts to find data useful to the task at hand in the absence of traditional relational schema definitions~\cite{nargesian2019tutorial, chepurko2020ardaautomaticrelationaldata, zhang2020, dong2021efficient}.
Features of existing approaches are summarized in Table~\ref{tbl:relatedwork}.
Most commonly, existing table search methods employ keyword search~\cite{auctus, brickley2019, google_dataset_search, liu2007tablerank}, where the task is to find tables with matching file names, captions, and metadata annotations to the user query.
\emph{However, the reliance on metadata assumed to be available is restrictive.}
As an alternative, other methods employ the query-by-example paradigm, where the user construct an example table as query~\cite{zhu2016lshensembleinternetscaledomain, nargesian2018table, rezig2021dice, Rafiei2022, bogatu2020, chepurko2020ardaautomaticrelationaldata, santos2022, fernandez2018}.
This allows for \textit{relevance}-based table search, where retrieved tables are not limited to only those containing exact matches.
As an example, BareTQL~\cite{Rafiei2022}, JOSIE~\cite{josie}, LSH Ensemble~\cite{zhu2016lshensembleinternetscaledomain}, and MATE~\cite{esmailoghli2022matemultiattributetableextraction} measure the overlap between column values, where \emph{a higher overlap is considered a signal of higher relevance}.
These approaches also permit semantic differences in the representation of values by using KG mappings and value embeddings~\cite{nargesian2018table}.
Yet, data integration approaches, like Proximity~\cite{alserafi2020keeping}, that focus on the specific task of table and schema matching are ill-suited to address data discovery tasks.

In dataset augmentation, the aim is to add features, labels, or instances to a dataset~\cite{zhang2020}.
ARDA~\cite{chepurko2020ardaautomaticrelationaldata} evaluates the quality of the obtained augmenting information.
ARDA works as a two-step algorithm: firstly, it searches for joins and then prunes out irrelevant features with feature selection algorithms.
It takes as input an input dataset, a data repository, and a predictive model, and it then outputs an augmented dataset that results in an improved performance by the predicted model.
However, \emph{these methods are limited to only data lake values} and consider only a limited degree of external information.

Several methods~\cite{santos, fernandez2018aurum, nargesian2018table, Rafiei2022, da2025} go beyond simple value overlap and use KG mappings and embeddings of values, tuples, or columns to handle heterogeneous representations of data values.
Some even exploit taxonomic KG information and KG properties to compute the relatedness of two sets of entities in table columns~\cite{nargesian2018table, santos}.
For web tables, some machine learning methods~\cite{balog2020, balog2021, turl} exploit contextual information in the web pages of the web tables to obtain a better high-level understanding of the table contents.
However, these methods heavily rely on contextual information and require training complex machine learning models.
Furthermore, some of these methods require that a query is compared to all data lake tables, which is infeasible in a scalable data lake system.

TURL~\cite{turl} is a representation learning method for table understanding.
Although it was not designed for table search, it vectorizes the query input and data lake tables, and hence, these vectors can be used for table search via vector similarities.
However, the data lake tables and input query must be sufficiently large for the vector representations to be of high quality.

Many systems perform union search, where the goal is to find unionable tables~\cite{nargesian2018table, santos, fan2023semanticsawaredatasetdiscoverydata, bogatu2020}.
Union search does not require syntactic overlap and can utilize semantic similarity to establish when pairs of columns describe the same domain of values.
However, this type of search is limited to only rank tables based on structural similarity to the query table.
Many other systems perform join, i.e., they find tables that are joinable with a query table~\cite{bogatu2020, josie, santos2022, zhu2016lshensembleinternetscaledomain, esmailoghli2022matemultiattributetableextraction, cafarella2009data, kokel2025evaluatingjoinablecolumndiscovery}.
These systems rank tables based on the syntactic overlap in the content of columns or rows, but they do not account for any notion of topical relevance or semantic similarity.
Instead, semantic search is not only interested in unionable and joinable tables but also considers the strength of semantic similarity.
Another approach exploits contextual data derived from static analysis of the usage of tables in different programs, such as Python notebooks~\cite{zhang2020}.
This approach is limited in the adopted type of contextual information, but it shows the importance of integrating contextual information in table search.

Rapidminer~\cite{lehmberg2017winte, lehmberg2017stitching, rapidminer} and S3D~\cite{Galhotra2020} better exploit the taxonomic information available in KGs by explicitly assuming that data lake tables contain only a set of entities (\emph{entity tables}), i.e., each row is considered an entity, where one column represents the entity identifier, and the remaining are entity attributes~\cite{DasSarma:2012}.
This allows them to first perform a form of schema matching between the table columns and the schema of the KG, and then use this information to retrieve a set of attributes to explain joins with other tables.
\emph{However, these methods do not fully exploit the semantic resources or are limited to explicit one-hop-matching with table attributes.}

Thetis~\cite{christensen2025fantastic} employs an entity-centric semantic relatedness measure for generic entity tuples without requiring matching any structural condition between tables.
Of all methods, only Thetis accepts generic \emph{entity tuples} and utilizes both \emph{content} and \emph{semantic relationships} to produce a ranking of tables based on relevance that goes beyond structural matching.
Furthermore, Thetis also retrieves tables that contain \emph{related} entities without requiring metadata and schema matching.

\subsection{Semantic Data Lakes}
Knowledge graphs are widely adopted across organizations~\cite{noy2019industry, saga} and have also been exploited in data lakes~\cite{ontario, squerall, constance, bernhauer2022open, bagozi2019, schmid2019using} such that the schema and contents of the datasets in a data lake are linked to the entities and relationships in a KG.
These links enable uniform retrieval of heterogeneous data lake datasets through semantic queries.
However, this approach requires significant cleaning efforts and manual mapping~\cite{sequeda2019pay}, despite achieving a high level of data integration.
Furthermore, these significant efforts contrast with the principle that a data lake should allow effortless addition of new datasets.

\emph{Thetis~\cite{christensen2025fantastic} represents instead a more flexible option}, as it automatically identifies all mentions of KG entities within each dataset.
As a result, Thetis obtains a set of datasets, each one containing mappings (also called links) to nodes in the KG.
To this end, Thetis is effective with only a partial mapping coverage of a dataset schema to KG relationships.
Therefore, the integration of a KG into a data lake makes Thetis a semantic data lake without manual curation.

Thetis~\cite{christensen2025fantastic} allows for a wide range of entity-centric similarity functions, e.g., cosine similarity of entity embeddings or Jaccard similarity of entity KG types, which is contrary to approaches like S3D~\cite{Galhotra2020}.

\subsection{Multi-Criteria Ranking in Data Lakes}
Multi-criteria ranking is a concept often used in recommender systems~\cite{zheng2022, Ceballos2016, puvska2024introducing}.
The aim is to combine ranking according to multiple criteria (i.e., multiple preference factors) into a single, overall ranking.
Thus, multi-criteria ranking can be exploited in table search systems to combine rankings from multiple search systems, which has recently been shown to improve precision and recall~\cite{kokel2025evaluatingjoinablecolumndiscovery}.
\emph{Yet, multi-criteria ranking has received limited attention and has not been explored in data lakes and when combining search methods that incorporate different types of relevance signals~\cite{eric2026, mohamed2006, khamnuansin-etal-2024-mrrank}.}

Thetis~\cite{christensen2025fantastic} has shown that complementing traditional keyword search with semantic search can significantly improve the recall of both search systems.

One approach~\cite{eric2026} performs multi-criteria ranking over text documents by combining keyword search with vectorized representations of the documents using Sentence-BERT~\cite{reimers-gurevych-2019-sentence}.
However, the approach does not incorporate any semantic information from KGs and is limited to the information present in the input data.
Another work~\cite{mohamed2006} combines different token-based measures (e.g., term frequency, inverse document frequency, and document length) using decision rules, which are cumbersome to define and set up.
Furthermore, for each input query, the approach applies the decision rules to every possible pair in the corpus to determine their ranking, requiring quadratic comparisons per query, thereby severely limiting the method's scalability.

TOPJoin~\cite{kokel2025topjoin} is a multi-criteria join search method incorporating table context.
It aims to retrieve tables with semantically joinable columns.
It uses three independent methods based on syntactic joinability, table metadata embeddings (e.g., table name and description), and value semantics, where columns are represented as embeddings of their values.
It combines relevance signals using TOPSIS~\cite{tzeng2011multiple} to obtain a single ranking using 6 different table features.
However, similar to other join search methods, TOPJoin also suffers from the join search limitations in the data lake setting.
Similarly, TACTUS~\cite{sun2026efficienteffectivetablecentrictable} proposes a table-centric approach for union search.
TACTUS does not rely on metadata and instead focuses on holistic table-level semantics, in which a single embedding represents each table.
These embeddings are generated using self-supervised learning.
Given a query table as input, it generates a single embedding representation and then performs nearest-neighbor search over table embeddings.
The retrieved candidates are combined by their table-level scores and column alignment scores to produce a final unionability score.

Blend~\cite{blend} unifies join, union, and correlation discovery.
It consists of a \textit{seeker} that performs join, keyword, union, and correlation search and then combines these results by intersection, union, difference, and frequency-based ranking.
However, this approach is still limited by the join and union search constraints, and it requires users to define rules specifying how the discovery should be carried out.

More recent approaches~\cite{khamnuansin-etal-2024-mrrank, Debabrata2023, Pobrotyn2020ContextAwareLT} apply learning-to-rank, which requires extensive training, often with supervision.
One method~\cite{khamnuansin-etal-2024-mrrank} requires training a neural retriever model to retrieve top-\textit{K} results.
For each of these results, it computes a BM25 score as well as additional neural retriever scores, which it uses to construct a feature vector.
They train a supervised learning-to-rank model and use the constructed feature vectors to re-rank the results retrieved by the neural retriever.
Thus, this approach, like others, is dependent on the contents of the documents only.
\section{Semantic Table Search} \label{sec:approach}
In this work, we address the need for retrieving tables from a data lake related to one or more entity tuples provided as a query.
We first provide the general problem formulation, then we formalize the notion of semantic relevance of a table given the semantic relevance score for entities, and finally, we provide an axiomatization of the properties that an appropriate semantic relevance score should satisfy.
The definition of a semantic relevance score of a table and for entities, as well as the axioms, are provided by Thetis~\cite{christensen2025fantastic}.

\subsection{Aggregating Semantic Relevance}
As defined above, semantic table search relies on the definition of a semantic relevance score, {\semrel}$(Q,T)$, between a query $Q$ and a table $T$.
Since the query is composed of one or more entity tuples, and the table is composed of one or more rows containing entity mentions, in practice, the relevance score is a function that determines whether the contents of a table are relevant to the query by comparing each entity tuple in the query with each row in the table.
Note also that, since the relevance score is measured through $\graph$, we only consider the entity mentions in a given table row, extracted by the mapping function $\mapfun$ (see Definition~\ref{def:semDL}).
This means that table rows are effectively treated as entity tuples.
Therefore, we define {\semrel} based on the definition of a \emph{semantic relevance score} between pairs of entity tuples ${t_i}, {t_j}{\in}\powerset(\nodes)$.
Given the set of all possible tuples as the set $\powerset(\nodes)$, consisting of all subsets of an arbitrary number nodes ${\nodes}$ from the graph {\graph}, we define the semantic relevance function as ${\semrel}{:}\powerset(\nodes){\times}\powerset(\nodes){\mapsto}{[0,1]}$.
For example, the pair of tuples $t_1$:\ituple{Truck Hannah, Los Angeles Angels} and $t_2$:\ituple{Fresco Thompson, Birmingham Barons} should be given a higher relevance score than the pair $t_1$:\ituple{Truck Hannah, Los Angeles Angels} and $t_3$:\ituple{Portsmouth Cubs}.
$\semrel$ can be asymmetric, as two tuples can contain a different number of entities.
Nonetheless, for consistency, with $t_3{\subset}t_1$ and $|t_3|{<}|t_1|$, it must always hold that $\semrel(t_1, t_3){\leq}\semrel(t_3, t_1)$.
Therefore, when $t_1$ is considered as a query, the relevance score must signal that $t_3$ only partially matches the type of information in $t_1$.
On the other hand, when $t_2$ is considered as a query, $t_1$ is a perfect match.

Defining the semantic relevance between two entity tuples $t_1, t_2$ requires $\semrel$ to compare each entity in $t_1$ with each entity in $t_2$.
Moreover, the semantic relevance between tuples relies on the notion of \emph{semantic similarity} between any two entities in $\graph$.
For example, in the simplest case, two tuples compared to each other consist of just a single entity, i.e., ${t_i}:$\etuple{e_i} and ${t_j}:$\etuple{e_j}.
Here, the highest relevance score is that of an entity compared to itself, where for non-identical entities, it is possible to determine the entity similarity through ${\graph}$.
Given a query tuple $t_Q$, a target tuple $t_T$ is relevant if it contains exactly that entity or an entity that is semantically similar for every entity in $t_Q$.
Hence, we require the definition of the concept of \emph{semantic similarity score} $\score{:}{\nodes}{\times}{\nodes}{\mapsto}{[0,1]}$, with $\score(e,e){=}1$.
The semantic similarity score between pairs of entities should satisfy all properties of a metric.

Given the semantic relevance score defined between pairs of tuples, he {\semrel} score between a query $Q$ and a table $T$ is defined as an aggregate score of the combined relevance for each tuple in the query, $t_i{\in}Q$ and each tuple in the target table, $t_j{\in}T$.
This is a requirement, as we must compute a single table relevance score such that we can produce a ranking of the tables to satisfy Problem~\ref{prob:combined_table_search}.
We instantiate {\semrel} as the average of the best match between query tuples and tuples in the table.
However, there can be different instantiations for {\semrel}.
We define {\semrel} as:

\begin{equation}
    \label{eq:max}
    {\semrel_{MAX}}(Q,T){:}\frac{\sum_{{t_i}{\in}Q}\max_{{t_j}{\in}T}{\semrel}({t_i},{t_j})}{|Q|}
\end{equation}

The problem of semantic table search, hence, is different from classical table search, e.g., used for table augmentation, because it does not necessarily require for the target tables to contain the entities appearing in the query to be relevant.

\subsection{Axiomatization of Relevance}\label{subsec:axiom}
We provide an axiomatization of the properties that an appropriate relevance score between entity tuples must satisfy.
We use these axioms to design our relevance score.

Given a tuple ${t_Q}{:}${\etuple{e^{1}_{Q},...,e^{m}_{Q}}} from the input query $Q$ (i.e., ${t_Q}{\in}Q$), and tuple ${t_T}{:}${\etuple{e^{1}_{T},...,e^{n}_{T}}} from the table $T{\in}\Dl$ (i.e., ${t_T}{\in}T$), the concept of a \emph{relevant mapping} from $t_Q$ to $t_T$ is defines as a partial injective function\footnote{We will write $\mu_{T}$ or $\mu$ instead of $\mu_{t_Q,t_T}$ to simplify notation.} $\mu_{t_Q,t_T}{:}t_Q{\hookrightarrow}t_T$ s.t. $\mu_{t_Q,t_T}(e^{i}_{Q}){=}e^{j}_{T}$ iff $\score(e^{i}_{Q},e^{j}_{T}){>}0$.
Therefore, there are the following four cases:

\begin{enumerate}[leftmargin=*]
    \item All entities in $t_Q$ appear (separately) in $t_T$, i.e., $\forall$$e^{i}_{Q}{\in}t_{Q}$ . $\mu(e^{i}_{Q}){\equiv}e^{i}_{Q}$, this is called a \emph{total exact mapping} denoted $t_{Q}{\tem}t_{T}$.
    \item Some, but not all, entities in $t_Q$ appear in $t_T$, i.e., ${\exists}{t'_{Q}}{\subset}t_{Q}$ s.t. $\forall$$e^{i}_{Q}{\in}t'_{Q}$ . $\mu(e^{i}_{Q}){\equiv}e^{i}_{Q}$, this is called a \emph{partial exact mapping} denoted $t_Q{\pem}t_T$.
    \item For each entity $e^{i}_{Q}{\in}{t_Q}$ there exist a mapping entity $e^{j}_{T}{\in}t_T$,
    i.e., $\forall{e^{i}_{Q}{\in}t_{Q}}$ . $\exists{e^{j}_{T}{\in}t_{T}}$ s.t. $\mu(e^{i}_{Q}){=}e^{j}_{t}$ and $\score(e^{i}_{Q},e^{j}_{T}){>}0$,
    those are called related entities. Since $\mu$ is injective, no two entities in $t_Q$ are mapped to the same entity in $t_T$, this is called a \emph{total related mapping} denoted $t_Q{\tam}t_T$.
    \item There exists a related mapping as defined above but just of a subset of the entities in $t_{Q}$, this is called a \emph{partial related mapping} denoted $t_Q{\pam}t_T$.
\end{enumerate}

{\noindent}As an example, consider the following tuples:
$t_1$:\ituple{Mitch Stetter, Milwaukee Brewers}, 
$t_2$:\ituple{Mitch Stetter, Milwaukee Brewers, Milwaukee},  
$t_3$:\ituple{Ron Santo, Chicago Cubs},  
$t_4$:\ituple{Ron Santo, Chicago},
$t_5$:\ituple{Milwaukee}.
The following holds: $t_1{\tem}t_2$, $t_2{\pem}t_1$, $t_1{\tam}t_3$, $t_2{\pam}t_4$, and
$t_1{\pam}t_5$.

When none of the above holds for a target tuple, that tuple $t_T$ is considered \emph{irrelevant} to the query tuple $t_Q$ and is excluded from the output.
Furthermore, note that the definition of total \emph{related mapping} also holds for the case in which some of the mapped entities in $t_T$ are exact mappings for the entities in $t_Q$.
Specifically, if a mapping exists that such that all entities in $t_Q$ are mapped (by exact or related mapping) to at least one entity in $t_T$, this case is considered a total related mapping (e.g., $t_4{\tam}t_2$ from above).
We identify a set of foundational axiom based on the above cases that the semantic relevance score must satisfy given the type of mapping between two tuples.
Namely, given a query tuple $t_{Q}$  and two distinct target tuples $t_{T1}$ and $t_{T2}$, the following must hold:

\sloppy
    \begin{axiom}\label{axiom1}
      if it holds that $t_{Q}{\tem}t_{T1}$ and $t_{Q}{\ntem}t_{T2}$ then it must also hold that ${\semrel}(t_{Q}, t_{T1})$ ${>}{\semrel}(t_{Q}, t_{T2})$, i.e., total exact mappings are the most relevant.
    \end{axiom}

    \begin{axiom}\label{axiom2}
    if it holds that $t_{Q}{\pem}t_{T1}$ and $t_{Q}{\pem}t_{T2}$$\vee$$t_{Q}{\pam}t_{T2}$ for the same set or a subset of entities in $t_{Q}$, i.e., $dom(\mu_{T2}){\subseteq}dom(\mu_{T1})$, then it must also hold that ${\semrel}(t_{Q}, t_{T1}) \geq {\semrel}(t_{Q}, t_{T2})$, i.e., larger partial exact mapping are more relevant then mappings that involve only a subset of the entities.
    \end{axiom}

    \begin{axiom}\label{axiom3}
    if it holds that ${\forall}{e^{i}}{\in}t_{Q}$  $\score(e^{i}, \mu_{T1}(e^{i}))$ ${>}\score(e^{i}, \mu_{T2}(e^{i}))$, then it must also hold that ${\semrel}(t_{Q}, t_{T1}) > {\semrel}(t_{Q}, t_{T2})$, i.e., tuples composed by more related entities should attain a higher relevance score.
    \end{axiom}

In the following, we define the $\semrel$ relevance score that satisfies all the above axioms and an algorithm to compute such a score. 
Later, in Section~\ref{sec:approximate}, we introduce an optimized algorithm able to compute an approximate solution with performance guarantees.
\section{Semantic Search Algorithm} \label{sec:exact}

\begin{figure*}[t]
    \includegraphics[width=\textwidth]{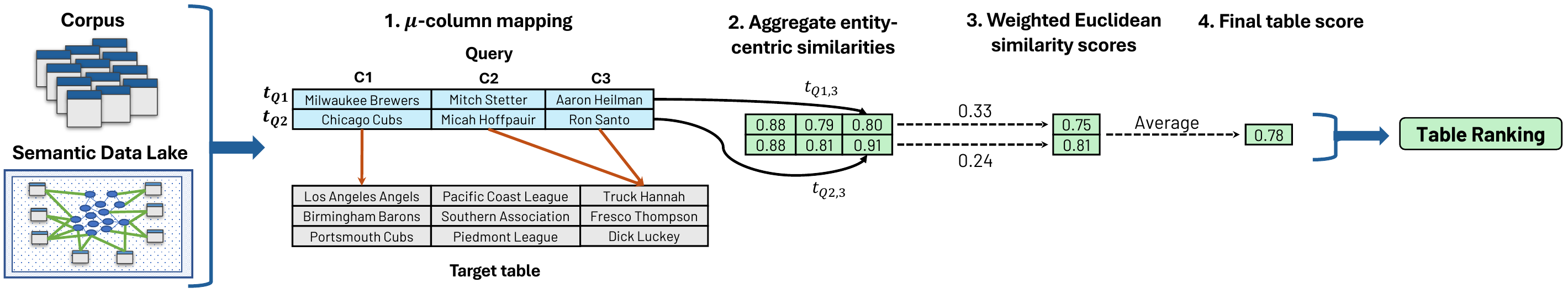}
    \vspace{-10pt}
    \caption{Semantic table search algorithm}
    \label{fig:algorithm}
\end{figure*}

We first describe how to produce a mapping between a query and the tuples in a table; then we describe the computation of the {\semrel} score for a pair of tuples.
Finally, we present and describe the exact algorithms solving semantic table search.
The algorithm is illustrated in Figure~\ref{fig:algorithm} and is based on Thetis~\cite{christensen2025fantastic}.

\subsection{Mapping the Query Tuple to a Table}
The first step of computing a relevance score between a query entity tuple $t_Q$ and a target entity tuple $t_T$ is to identify a mapping between the two.
Ideally, based on the semantic relevance score $\score$, we define a mapping $\mu_{Q,T}$ such that we maximize the cumulative score $\sum_{e_i\in t_Q}\sigma(e_i, \mu_{Q,T}(e_i))$.
Since the semantic relevance score must be aggregated over all tuples of the target table, given that two tuples ${t}^{1}_{T}$ and ${t}^{2}_{T}$ are from table $T$, and we map entity $e_i$ from ${t}_{Q}$ to entity $e_j$ from ${t}^{1}_{T}$, corresponding to attribute $A{\in}\At$, then, when comparing ${t}_{Q}$ to ${t}^{2}_{T}$, we must also map $e_i$ to the entity $e_l$ corresponding to $A$ for ${t}^{2}_{T}$.
Specifically, we map n entity from the query to entities from the same column for all tuples in the target table.
Hence, given a semantic similarity score $\score$, a query tuple $t_Q$, and a target table $T$, for each entity in the tuple, we need a mapping to a column in a table that maximizes the total score $\semrel$ for all entities in that column.
We must furthermore ensure that every query tuple entity is assigned to a unique target table column.

Let a table $T$ contain cell values organized across columns $C_1,\dots,C_n$  and rows $R_1,\dots,R_m$.
Furthermore, let $t_Q$ denote a query tuple consisting of entities $e_1,\dots,e_k$.
Then, we define a column-relevance score between a query entity $e_i{\in}t_Q$, and column $C_j{\in}T$ as:

$$score(e_i, C_j) = \sum_{\bar{e}\in C_j}\score(e_i, \bar{e})$$

{\noindent}Since we want the final relevance score, summed across all query entities, to be maximized, we compute the relevance score for each pair of query entity and table column to find the best mapping function $\tau{:}t_Q{\mapsto}\{C_1, ..., C_n\}$.
The mapping score from query entities to target table columns is represented in a matrix, $S$, as follows:

$$S =\begin{pmatrix}
score(e_1,C_1) & \cdots & score(e_1,C_n)\\
\vdots & \ddots& \vdots\\
score(e_k,C_1) & \cdots & score(e_k,C_n)\\
\end{pmatrix}$$

{\noindent}Our goal is to identify an assignment for each query entity to a unique target table column such that column relevance score is maximized, where each entity must map to a different target table column.
This statement is formalized as the following optimization problem:

$$\argmax_X \sum_{i=1}^{k}\sum_{j=1}^{n}S_{ij}X_{ij}$$

where the assignment is specified by $X$ which is a boolean matrix, where $X_{ij}{=}1$ if and only if row $i$ is assigned to column $j$, and where each row is assigned to exactly one column, and each column is assigned to at most one row (i.e., in the matrix $X$, there are exactly $k$ ones).
We solve this assignment using the Hungarian Method~\cite{EdmondsK72}.
We can express this as permuting the rows and columns of a cost matrix C to maximize its trace $\min_{L,R}{{Tr}}(LCR)$ where L and R are permutation matrices.
The solution to the assignment problem leads to the mapping function $\tau_{Q,T}{:}t_Q{\mapsto}\{C_1, ..., C_n\}$.
Hence, given the query tuple $t_Q{\in}Q$ and target row $t_T{\in}T$, to compute $\mu_{Q,T}$, we map each entity $e_j{\in}t_Q$ to the entity $e_k{\in}t_T$ where $e_k{=}C_k(t_T)$, i.e., the entity corresponding to column $C_k$ in $t_T$, with $C_k{=}\tau(e_j)$.

The result of $\mu_{Q,T}$ is independently computed for each query tuple.
Therefore, it is not required that all query tuples $t_Q{\in}Q$ follow the same schema.
This process is depicted in Region 1 of Figure~\ref{fig:algorithm} (for simplicity, both query entity tuples have the same column mapping).

\subsection{The definition of the {\semrel} score}
Here, we provide a definition for the similarity measure that satisfies all the properties required by the {\semrel} score.
Given the query tuple ${t_Q}{:}${\etuple{e^{1}_{Q},...,e^{m}_{Q}}} (with $|t_Q|{=}m$) from the input query $Q$ (i.e., ${t_Q}{\in}Q$), and the target tuple ${t_T}{:}${\etuple{e^{1}_{T},...,e^{n}_{T}}} from the table $T{\in}\Dl$, we map the target tuple to an Euclidean space, where the number of dimensions is equal to the number of query tuple entities, i.e., ${\R}^{m}$.
Given a relevant mapping $\mu{:}{t_Q}{\mapsto}{t_T}$, each target tuple $t_T$ is mapped to a point $p_T{:}{\langle}x_1,...,x_m{\rangle}$ with coordinate $x_i{=}{\score}(e^{i}_{Q}, \mu(e^{i}_{Q}))$ in the Euclidean space.
In case there is no relevant mapping in $t_T$ for an entity $e^{i}_{Q}{\in}t_{Q}$, then, from the definition of relevant mapping, it follows that $x_i{=}0$.
This is depicted in Region 2 of Figure~\ref{fig:algorithm}.
Given each tuple mapping the Euclidean space, we compute the semantic relevance of each tuple as their \emph{euclidean distance} $\textsc{D}$ from the perfect match, which is from the point corresponding to $t_Q$ with coordinates $x_i{=}1$ $i{\in}[1,m]$.
This is illustrated in Region 3 of Figure~\ref{fig:algorithm}.

Tuples containing all the user query entities are the most relevant, i.e., those for which it holds $t_Q{\tem}t_T$ (Axiom~\ref{axiom1}).
These tuples have scores $\score(e^{i}_{Q}, e^{j}_{T})=1$ for $e^{j}_{T}{=}\mu(e^{i}_{Q})$.
Secondly, combining axiom 2 and axiom 3, we derive that apart from total exact mappings, the second best set of tuples are those that are total related mapping, but where most of the tuples in $t_T$ are exact mapping of tuples in in $t_Q$, i.e., when it holds that $t_Q{\tam}t_T{\wedge}t_Q{\pem}t_T$.

According to the definition above, all tuple entities contribute contribute to the final relevance score in the same way.
Although, the query entities have different roles, as some entities are more important in determining the relevance of a given target tuple.
For instance, given a query \ituple{Mitch Stetter, Milwaukee Brewers}, we can intuitively assume that the user's interest is primarily about baseball players and their team to a smaller degree.
Thus, a tuple \ituple{Milwaukee Brewers} is intuitively less relevant than the tuple \ituple{Mitch Stetter}.
Therefore, to handle this, we propose an instantiation of ${\semrel}$ that is able to capture the need for differentiating the discriminate power of the entities in the user query.
We thereby introduce the concept of informativeness of an entity, i.e., $\Inf{:}\nodes{\mapsto}[0,1]$ that automatically assigns a weight to the query entities based on entity frequency in the corpus.
We compute {\semrel} as a the \emph{weighted euclidean distance} as follows:

\begin{equation}
    \textsc{D}_{\Inf}(p_Q,p_T)=\sqrt{\sum_{i{\in}[1,m]} {\Inf{(e^{i}_{Q})}}\left(1-x_{i}\right)^{2}}
\end{equation}

{\noindent}Finally, we convert the \emph{weighted Euclidean distance} into a similarity score, such that a score closer to 1 means smaller Euclidean distance.

\begin{equation}
    {\semrel}(t_Q,t_T)=\frac{1}{\textsc{D}_{\Inf}(p_Q,p_T)+1}
\end{equation}

The final table score is the average of the weighted euclidean distances (illustrated in Region 4 of Figure~\ref{fig:algorithm}).

\subsection{Computing Semantic Relevance} \label{subsec:algo}

\begin{algorithm}[t]
\caption{Thetis Semantic Table Search}
\label{alg:generator}
\KwIn{${\Dl}{=}\{{T_1}, {T_2}, {...}, {T_n}\}$; $Q{:}\{{tq_{1}}, {tq_{2}}, {...,} {tq_{m}}\}$}
\KwIn{$\Inf{:}\nodes{\mapsto}[0,1]$; ${\score}{:}\nodes{\times}\nodes{\mapsto}[0,1]$}
\KwOut{\emph{Table Relevance scores} ${\langle}T_i, {\alpha}_{i}{\rangle} \forall T_i{\in}\Dl$}

tableScores ${\gets}$ [ ]\;
\ForAll{$T{\in}{\Dl}$}{
    qScores ${\gets}$ [ ]\;
    \ForAll{$tq{\in}{Q}$}{
        columnMapping${\gets}${hungarianMapping($tq$, $T$)}\;
        rScores${\gets}$ [ ]\;
        \ForAll{\emph{row}${\in}{T}$}{
            
            eScores${\gets}$ [ ]\;
            \ForAll{\emph{e}${\in}{tq}$}{
                mappedToColumn${\gets}$columnMapping[$e$]\;
                eScores$[e]{\gets}\score(e, \text{row[mappedToColumn]})$\;
            }
            rScores$[\text{row}]{\gets}${eScores}\;
        }
        aggScores${\gets}${aggRowScores(rScores)}\;
        qScores$[tq]{\gets}$$1/(\sqrt{\sum_{e{\in}tq} {{\Inf}(e)}\left(1-\text{aggScores}[e]\right)^{2}}+1)$\;
    }
    tableScores $[T]{\gets} (\sum_{tq{\in}Q}$qScores$[tq])/|Q|$\;
}

\end{algorithm}

The procedure described by Algorithm~\ref{alg:generator} computes the final semantic relevance score of all data lake tables.
The input to the algorithm is set of tables $T_i{\in}{\Dl}$, a query $Q$ of a set of tuples $t_{Q_i}{\in}Q$, an entity weighing function $I$, and an entity semantic similarity scoring function ${\score}$.
The output is the semantic relevance score {\semrel} for each data lake table.

The implementation of Thetis~\cite{christensen2025fantastic} currently supports two alternative similarity functions for entities, but can be extended to support more.
The first is the Jaccard similarity of the sets of entity types of two entities, and the second is the cosine similarity between pairs of entity embeddings.
For the first similarity function, two entities are considered similar if they share the same set of entity types.
In rich KGs, entities are commonly annotated with with multiple types at different granularity levels (e.g., \texttt{Milwaukee Brewers} is annotated both as a sports team and as an organization in DBpedia).
Therefore, we adjust the Jaccard similarity score such that an entity compared to itself is assigned a similarity of 1.0, while we cap the Jaccard similarity at $0.95$ otherwise.
We therefore define our adjusted Jaccard similarity given $\mathcal{T}_1$ as the set of entity types for entity $e_i$ as:

\begin{equation}
Jaccard\text{*}(e_1, e_2)= \begin{cases}1 & \text { if } e_1{\equiv}e_2\\ 
                   \min(0.95, \frac{|\mathcal{T}_1 \cap \mathcal{T}_2|}{|\mathcal{T}_1 \cup \mathcal{T}_2|}) & \text { otherwise }\end{cases}
\end{equation}

Entity embeddings are obtained through a self-supervised machine learning approach that learns a vector representation for every entity in the KG based on their higher-order connections to other entities~\cite{EntityType2022,ristoski2016rdf2vec}.
Therefore, two entity embeddings are similar if the two corresponding entities have similar semantic connections within the graph.
Given the entity embedding vector $\mathbf{V}_i$ for $e_i$, the semantic similarity ${\score}(e_1,e_2)$ is computed as the cosine similarity between their vectors:

\begin{equation}
    Cosine(e_1, e_2) = \frac{\mathbf{V}_1 \cdot \mathbf{V}_2}{\lVert \mathbf{V}_1 \rVert \lVert \mathbf{V}_2 \rVert}
\end{equation}

\emph{Note that this Thetis search framework and the optimization algorithm are designed to generalize to other similarity scores}.
For example, the Jaccard similarity can also be computed for two sets of entity predicates in the KG~\cite{mottin2016exemplar} and the RDF2Vec entity embeddings can be replaced with other entity embedding approaches designed specifically for node classification~\cite{EntityType2022, accelerating_linking}.

We now describe Algorithm~\ref{alg:generator} for computing a relevance score for all data lake tables.
Given a data lake table, we compute a mapping $\tau{:}t_Q{\mapsto}\{C_1, ..., C_n\}$ from each query entity to a table column using the Hungarian Method to maximize the summed similarity score across all query entities (line 3).
We then compute a mapping and relevance score from each query tuple to each table row (lines 7-12), which involves computing the semantic similarity score between each entity in the query tuple $e{\in}t_Q$ and each entity in the target table row $e'{\in}t_T$.
We aggregate every row score per query tuple by maximum (line 13), such that is extracts the maximal semantic similarity score among all entities mapped to the same query entity.
Given an aggregated tuple relevance score for $t_Q$, we compute the Euclidean distance from $t_Q$ to the point $x_{i}=1{\text{ }}i{\in}[1,|t_Q|]$ which is converted to a distance similarity score $0{\leq}s{\leq}1$, where $s$ closer to 1 means higher similarity (line 14), which allows weighing the similarity based on the relative frequency of query entities in the corpus.
Finally, we aggregate these tuple scores by average (line 15).

The time complexity of Algorithm~\ref{alg:generator} is determined by the mapping function $\mu_{T,Q}$.
Given the number of table rows $R_T$, the number of table columns $C_T$, the number of query tuples $R_Q$, and the number of query entities per query tuple $C_Q$, the time complexity is defined as $\mathcal{O}(R_{T}C_{T}R_{Q}C_{Q})$.
Although, the query is usually small, the time complexity is in practice $\mathcal{O}(R_{T}C_{T})$.
\section{Combined Table Search} \label{sec:combined_search}
We now describe our proposed approach to combine rankings of BM25 keyword search and Thetis semantic search.
Multiple approaches to combining ranking exist, especially within recommender systems.
We design an approach that combines two popular techniques, namely Pareto~\cite{Luc2008, yong2023} and Technique for Order of Preference by Similarity to Ideal Solution (TOPSIS)~\cite{tzeng2011multiple}.
Note that keyword search is likely to retrieve matching tables with many matching keywords, but without any relevant entities, e.g., searching for "\texttt{The Republic of the Congo}" might retrieve results containing The Republican Party due to the distinctive overlapping keyword.
Therefore, we exploit first Pareto optimality, as a high relevance score for a table in one approach does not compensate for a low score in another, i.e., a high relevance score from keyword search and a low score in semantic search will overall not lead to a high combined score.
Pareto does not compute a ranking, but rather fronts.
Hence, we apply TOPSIS to compute a combined relevance score within each front and to order tables within each front.
This ensures that only tables with high relevance scores from both keyword and semantic search retain a high ranking position.

Thetis produces relevance scores within $[0, 1]$, however, BM25 keyword search scores can be infinitely large.
We are therefore required to normalize these scores such that the relevance scores produced by both methods are aligned.
Based on our observation, log-scaled normalization provided the best performance.
Given a set of relevance scores for a single query $S:\{s_1, ..., s_k\}$, the log-scaled normalization for $s_i{\in}S$ is defined as:

\begin{equation}
    s_{i}' = \frac{log(1 + s_i)}{log(1 + \max S)}
\end{equation}

Both Pareto and TOPSIS are provided a 2-column matrix of relevance scores as input.
Rows are tables and columns are search methods, i.e., semantic and keyword search.
Given the ranking of each method, we perform a full outer join of the two result sets to construct the matrix, such that a table ranked by one method but not by the other is assigned a relevance score of 0.0 for the other method.

\subsection{Pareto Optimality}
Given a matrix $R$ of relevance scores, where rows are tables and columns are search methods, Pareto computes ordered fronts based on the notion of dominance, such that highest-scoring tables are assigned the highest ranking front.
Specifically, a table $T_i$ with relevance scores $S_i$ is said to dominate a table $T_j$ with relevance scores $S_j$ if the following properties hold:

\begin{enumerate}
    \item $\forall s_{i}{\in}S_{i}, s_{j}{\in}S_{j} : s_{i} \geq s_{j}$, where $i = j$
    \item $\exists s_{i}{\in}S_{i}{\land}\forall s_{j}{\in}S_{j} : s_{i} > s_{j}$
\end{enumerate}

{\noindent}To ensure that the Pareto fronts are large enough, we relax the above properties such that two scores are considered equal if their difference is less than $0.1$ (the value is chosen based on experimental evaluation).

Pareto does not compute a combined relevance score for each table, and hence, the tables within each front are unordered.

\subsection{TOPSIS}
We provide each Pareto front individually as input to TOPSIS, which computes a ranking of the tables contained within each front.
A Pareto front is a sub-matrix $R'{\subset}R$.
The idea of TOPSIS is to first compute the distance to the ideal and worst score for each table in the front, i.e., 1 and 0, respectively.
Then, we compute the relative distance to the ideal solution.
Given a table $T_i{\in}R'$ with relevance scores $S_i$, we compute the ideal and worst scores, $s^{+}_{i}$ and $s^{-}_{i}$:

\begin{equation}
    s^{+}_{i} = \sqrt{\sum_{s_{i}{\in}S_{i}} (s_{i} - 1)^2}
\end{equation}

\begin{equation}
    s^{-}_{i} = \sqrt{\sum_{s_{i}{\in}S_{i}} s_{i}^2}
\end{equation}

{\noindent}We then compute the relative distance to the ideal solution as:

\begin{equation}
    d_{i} = \frac{s^{-}_{i}}{s^{+}_{i} + s^{-}_{i}}
\end{equation}
\section{Semantic Search Prefiltering} \label{sec:approximate}
The approach described above computes the relevance score for all tables through an exhaustive brute-force search, an approach that is not scalable.
Therefore, here we propose a prefiltering approach so that we compare the query only against those tables that have a higher chance of being relevant.
As we show in our experiments, this results in a significant search space reduction whilst retaining the relevant target tables, which significantly improves the runtime of our table search algorithm.

To achieve this, we study both a Locality-Sensitive Hashing (LSH) scheme and Hierarchical Navigable Small World (HNSW)~\cite{hnsw, faiss} index.
The LSH index groups similar KG entities into the same buckets, whereas HNSW allows for approximate nearest neighbor search over entities.
Before executing our table search algorithm (Algorithm~\ref{alg:generator}), we search our LSH or HNSW index using the input query entities to reduce the search space by prefiltering tables.
Hence, all tables containing entities selected in this step are merged into the new, reduced search space.

\subsection{Building Locality-Sensitive Indexes}
We use the set of entity types or entity embeddings as input to LSH such that similar entities are hashed into the same buckets.
LSH takes as input a vector representation for the entity to compute a reduced signature for that vector, which is further divided into bands.
Each band is then hashed to find the corresponding bucket for similar entities; hence, each distinct band corresponds to a distinct group of buckets, and within each group of buckets, an entity appears only in one of them.

\noindent
\textbf{LSH for Entity Types.}
When comparing entities based on their types, we represent each type in the KG by a numeric index.
We mimic the idea of shingling in min-hashing for documents by creating a bit vector of size $|\mathcal{T}|{\times}|\mathcal{T}|$ for an entity, where $|\mathcal{T}|$ is the number of types, and flip bits to 1s in positions corresponding to pairs of types, e.g., a pair of types with indices 24 and 48 have index 2448 in the bit vector.
We hash this bit vector by LSH into an entity signature with a dimension equal to the number of permutation vectors.
Permutation vectors are a list of vectors containing indexes to permute the entity bit vectors.
Each permutation value $p_t$ in a permutation vector is a unique index in the range $0 \leq p_t < |T|$, where $|T|$ is the size of the type universe.
After computing an entity signature, the signature is hashed, and the entity is inserted into the corresponding bucket along a link to the tables in which the entity exists.
Some entity types are not informative due to their high frequency in the KG, such as the type \textit{owl:Thing} in DBpedia.
We therefore filter away those types that appear in more than 50\% of all tables in the corpus.
This percentage is chosen based on observations from smaller experiments.
Decreasing this percentage leads to a decrease in prefiltering efficacy.

\noindent
\textbf{LSH for Entity Embeddings.}
LSH schemes for embeddings are based on the concept of \emph{random projections}~\cite{faiss}.
Therefore, the signature dimension equals the number of projection vectors, each dividing the space into a positive and negative sub-space.
Then, the entity signature is a bit vector, where each 1 means that the dot product between the entity embedding vector and a given projection vector is positive.

\noindent
\textbf{LSH Index Structure.}
An LSH index has several groups of buckets. 
When inserting an entity, we split the signature into multiple bands: one for each group of buckets.
Thus, each band of the entity signature is hashed individually, and the entity is inserted into the corresponding bucket for that band.
It follows that an entity will appear in one bucket for each band.
The size of the bands and the number of permutation/projection vectors determine the number of bucket groups.
For example, an LSH index using 32 permutation/projection vectors and a band size of 8 has 4 bands and thus 4 bucket groups.
The number of buckets in each bucket group is $2^B$, where $B$ is the band size.
Therefore, large band sizes result in a large number of buckets, each more likely to contain only a few entities.
This corresponds to a larger search space reduction and smaller intermediate result sets to be ranked.
However, a higher search space reduction also risks loss of accuracy.
For each entity in the LSH index, we maintain a list of all the table identifiers in which that entity appears.

\subsection{Locality-Sensitive Hashing Index Prefiltering}
Searching an LSH index for each entity in the query is performed by first computing a signature for the entity.
The signature is split into multiple bands, and a bucket is found in each bucket group using the hash of the corresponding band.
When searching in the LSH index with a single entity, all entities in each matching bucket are merged into one set of entities, and the tables these entities are linked to are returned.
When searching in the LSH index with a single entity, all entities in each matching bucket are merged into one set of entities, and the tables these entities are linked to are returned.
Some tables may be found multiple times when merging the found buckets of entities and their linked tables. 
This gives the opportunity to further restrict the tables returned for a single entity by implementing a voting strategy.
That is, we treat this intermediate result set as a bag of tables, i.e., maintaining duplicates, and count the number of occurrences of a table in the results set, so that only those tables that appear a certain number of times are returned.
In our experiments, we will identify the ideal number of votes that achieves the highest positive impact on semantic table search.

\noindent
\textbf{Column aggregation.}
An alternative approach to compute entity signatures that also saves space is by aggregating vector representations of entities in the same column into one single vector representation. 
For this LSH index using types, we merge all entity types from the entities in a single table column into one unique set of types and compute a signature of that column using this merged set of types.
For embeddings, we compute the average embedding of all entities in a column.

Note that the higher the number of query entities, the higher the number of LSH lookups, which results in a larger result set.
We therefore further optimize the LSH lookup by applying aggregation by column on the input query level in the same way as described above for tables.
This introduces a further approximation with the benefit of reducing the search cost since it effectively treats queries composed of multiple entity tuples as if they were 1-tuple queries.

\subsection{Building HNSW Indexes}
For the case of dense vector search, HNSW indexes have shown higher response speed than LSH.
Therefore, we propose to replace the LSH index with an HNSW index for the case of semantic search based on entity embeddings.
In short, HNSW consists of multiple levels, each containing a graph larger than the one immediately above~\cite{hnsw}.
Therefore, we provide as input KG entities and their embeddings and build the HNSW index, such that similar entities are closely connected in the HNSW graphs.
When inserting an entity and its embedding, we compute the insertion level, which represents the first and highest level in which to insert the entity.
This insertion level is randomly assigned with a distribution following an exponentially decaying probability, such that higher levels contain fewer nodes.
Then, we start traversing from an entry node in the highest level and move along edges that connect to a node closer to the insertion node than the current node according to a vector similarity function (in our case, we use the cosine similarity).
Once we reach a local minimum, we move down to the next level and continue traversing from the local minimum in the previous layer.

Once we reach the insertion level, while we traverse the graph, we maintain a priority queue of $ef$ nearest neighbors that we have come across on this layer.
Once the local minimum has been reached, the priority queue contains the candidates to be connected to the insertion node.
A heuristic function selects $M$ of the $ef$ candidates and connects these to the insertion node such that we retain high global connectivity and high recall, while respecting that each node may have at most $M_{max}$ connections.
This process is repeated until reaching the lowest level, i.e., the base level.
In the base level, the process is the same, but each node is now allowed to have $M_{max0} > M_{max}$ connections.

\subsection{HNSW Search Space Prefiltering}
During search space prefiltering, we search the HNSW index by navigating the graphs to retrieve the nearest neighbors of a query vector representing a KG entity and the tables in which they exist.
Specifically, each level is traversed in the same way as during index construction.
Once the base level has been retrieved, the nearest neighbors are returned.
Every query entity is individually used to search the HNSW index, and the resulting set of nearest neighbor entities is merged across all query entities.
The tables containing any retrieved entity form the reduced search space.
\section{Evaluation} \label{sec:evalution}
We evaluate \texttt{Thetis++} in terms of retrieval effectiveness, runtime efficiency, scalability, and robustness. 
First, we study LSH and HNSW prefiltering techniques for semantic table search and identify configurations that maximize search-space reduction while preserving ranking quality. 
Next, we compare semantic and keyword-based table search and analyze the extent to which they retrieve complementary sets of relevant tables. 
Motivated by this complementarity, we evaluate \texttt{Thetis++} and several alternative approaches for combining semantic and keyword signals, including multi-criteria ranking, classifier-based method selection, and learning-to-rank. 
Finally, we assess scalability on large data lakes and investigate the impact of reduced entity-linking coverage.

\setlength{\abovecaptionskip}{2mm}
\begin{table}[!t]
\centering
\caption{Benchmark statistics: \#\, of tables (T), mean \#\, of rows (R), mean \# of columns (C), and mean entity link coverage (Cov)}
\label{tab:benchmark}
\adjustbox{width=\columnwidth}{
\begin{tabular}{r|r|r|r|r|r|r}
    \multirow{2}{*}{} &
        \multicolumn{2}{c|}{\textbf{Queries}} & \multicolumn{4}{c}{\textbf{Data Lake Tables}} \\
        & \textbf{T} & \textbf{C} & \textbf{T} & \textbf{R} & \textbf{C} & \textbf{Cov} \\ \hline
\textbf{WT 2015} & 100 & 3.4 & 238,038 & 35.1 & 5.8 & 27.7\% \\ \hline
\textbf{WT 2019} & 100 & 2.4 & 457,714 & 23.9 & 6.3 & 18.2\% \\ \hline
\textbf{GitTables} & 100 & 3.4 & 864,478 & 142.0 & 12.0 & 29.6\% \\ \hline
\hline
\textbf{Synthetic} & 100 & 3.4 & 1,732,328 & 9.6 & 5.8 & 34.8\% \\ \hline
\end{tabular}}
\end{table}

\subsection{Experimental Setup}
We implement keyword search in the well-established BM25~\cite{robertson2009probabilistic} algorithm.
We further compare to state-of-the-art unionability and joinability approaches SANTOS~\cite{santos}, Starmie~\cite{fan2023semanticsawaredatasetdiscoverydata} and D$^3$L~\cite{bogatu2020}, respectively.
We also compare against a deep-learning Table Representation, TURL~\cite{turl}, by adapting it for our task.
Using TURL's pre-trained model, we aggregate all contextualized vector representations in each table to construct an embedding for each table and query.
Cosine similarity between the table and query representations is used to rank the table search output.
We do not compare to other methods like Aurum~\cite{fernandez2018aurum} which rely on value equality between similar attributes, something already covered by BM25, or heuristics that have already been proven less effective than D$^3$L and SANTOS.
We also do not compare to SemProp~\cite{fernandez2018}, as it is an extension to Aurum, and its repository is outdated and non-operational.
Furthermore, we also do not compare against S3D~\cite{Galhotra2020}, as the code is not publicly available and RapidMiner~\cite{lehmberg2017winte,lehmberg2017stitching,rapidminer} because the code is not working, and the authors could not offer any help.

\texttt{Thetis++} with HNSW prefiltering and the evaluation suite are available as open-source\footnote{\url{https://github.com/dkw-aau/Thetis-extended}}.
Thetis~\cite{christensen2025fantastic} with LSH prefiltering is also available as open-source\footnote{\url{https://github.com/EDAO-Project/TableSearch}}.
We use a snapshot of DBpedia from 2021 as our reference KG containing $\sim$31M nodes, $\sim$89M edges, 763 distinct types, and 10,051 distinct predicates.
Evaluating \texttt{Thetis++} on other public KGs is beyond the scope of this work.
Nonetheless, a typical alternative would be WikiData, which showcases a rich vocabulary of entity and relationship types that is slightly richer than DBpedia's, and with which we would expect a slightly higher accuracy overall.
We generate RDF2Vec~\cite{ristoski2016rdf2vec} embeddings on our reference KG (which takes ~35 hours) to enable performing semantic table search using learned embeddings.
However, \texttt{Thetis++} can accommodate any set of entity embeddings.
\texttt{Thetis++} using entity types and embeddings perform similarly, and we will in our experiments focus on \texttt{Thetis++} using embeddings, unless otherwise specified.
We share both the code to generate these embeddings\footnote{\url{https://github.com/EDAO-Project/DBpediaEmbedding}} and the set of embeddings~\cite{RDFVecZenodo}.
Experiments are conducted on a server with 2TB of RAM and a 64-core CPU.

\spara{Alternative Multi-Criteria Ranking Techniques.}
In addition to \texttt{Thetis++}, we instantiate alternative techniques for multi-criteria ranking and study these.
We furthermore study variations of \texttt{Thetis++}.
First, we study a version of \texttt{Thetis++} that only uses TOPSIS for combining the rankings of Thetis and BM25, i.e., we skip the step using Pareto.
We call this \texttt{TOPSIS}.
Then, we study a version that aims to achieve high recall by first selecting the overlapping tables from the result sets of Thetis and BM25, referred to as $T$, and then filling out the remaining slots in the top-\textit{K} result set (i.e., $K-T$) with $\frac{K-T}{2}$ highest-ranking tables from Thetis and BM25 each.
We then re-rank this result set using either Thetis or TOPSIS, which we refer to as \texttt{Overlap-Thetis} and \texttt{Overlap-TOPSIS}, respectively.

We also study a classification-based version that predicts whether to use Thetis or BM25 based on the input query entities.
Based on our observation, queries containing entities that appear more frequently in the data lake and have more popular entity types in the reference KG tend to perform better in BM25.
Therefore, we train an XGBoost\footnote{\url{https://xgboost.ai/}} classifier that predicts which method to use.
To train this classifier, we represent each query entity by three numbers: (1) entity frequency, (2) maximum entity type frequency in the KG, and (3) minimum entity type frequency in the KG.
Then, a query is represented as an array, where each element contains the representation of a query entity.
We evaluate 1K 1-tuple and 5-tuple queries each in Thetis and BM25, and we use the label for the classifier as the search method that achieved the highest NDCG.
We refer to this method as \texttt{Classifier - frequency}.
We create an alternative to this classifier based on embeddings.
Given the RDF2Vec embeddings for KG entities, we compute the sum of the embeddings across all entities in a tuple.
We then compute the average embedding for each query tuple and let this be the representation of a query using the same label definition from \texttt{Classifier - frequency}.
We refer to this method as \texttt{Classifier - embeddings}.

Finally, learning-to-rank has recently been employed for combining multiple relevance signals to predict relevance scores for target tables~\cite{khamnuansin-etal-2024-mrrank, Debabrata2023, Pobrotyn2020ContextAwareLT}, i.e., we provide semantic relevance signals from entity embeddings as well as additional standard information retrieval features as input and predict a relevance score for each keyword-search-retrieved table.
We evaluate a learning-to-rank method~\cite{Pobrotyn2020ContextAwareLT} with these features, which we refer to as \texttt{Learning-to-rank}.
Specifically, for training, we provide as features content-based features consisting of the cosine similarity of the embedding representations of the query table and target table, where the embedding is computed the same way as in \texttt{Classifier - embeddings}.
Then, we provide as additional features the target table size (number of rows and columns) and the entity type overlap (computed as the Jaccard similarity) between the query entities and the target table entities.
The label is the ground truth relevance of the target table for the input query.
During query execution, we exploit the speed and accuracy of BM25 and therefore search with BM25 to retrieve a set of top-\textit{K} target tables, which we rank using the learning-to-rank approach.
BM25 is much faster than any other approach and is therefore chosen to construct this set of tables to be ranked given that BM25 candidate generation provides sufficiently high recall for learning-to-rank.

\spara{Data Lake Benchmarks.}
We evaluate \texttt{Thetis++} in an evaluation suite~\cite{stsd} containing real-world corpora based on Wikipedia tables from 2015 and 2019.
The suite comes with entity tuple queries consisting of various numbers of tuples (see characteristics in Table~\ref{tab:benchmark}).
We select a heterogeneous set of 100 1-tuple and 5-tuple queries from the set of queries in the evaluation suite of width of at least 3, where the 1-tuple queries are contained in the 5-tuple queries.
Note that, of the eighteen approaches mentioned in Table~\ref{tbl:relatedwork}, 5 of them evaluate on less than 50K tables and 8 on less than 500K tables.
The suite comes with entity links for table entity mentions; however, the entity linking quality is orthogonal to the ranking performance of \texttt{Thetis++}.
However, there are already numerous well-established entity linkers~\cite{accelerating_linking, lexma, bbw}.
The suite comes with ground truth rankings for each query based on Wikipedia categories and navigational links.
We use Normalized Discounted Cumulative Gain (NDCG) to evaluate the ranking performance and recall to evaluate the correctness of the returned result set.
We compute recall as the number of retrieved tables that are in the top-$K$ ground truth relevant tables.
In our experiments, unless specified otherwise, we focus first on results over WikiTables from 2015, which given the smaller size and the highest coverage, i.e., the number of cells linked to a KG entity, allows us to test more efficiently different settings.

For BM25, we convert the entity tuple queries to text queries by extracting the entire text contents from each row in the WikiTable from which the tuple originates.

To test scalability, we experiment with the GitTables dataset, since it contains 864,478 tables with an average number of columns and rows much larger than WikiTables (Table~\ref{tab:benchmark}).
Note that GitTables does not come with ground truth relevance annotations; hence, we cannot use GitTables to evaluate NDCG or recall.
We additionally generate a synthetic dataset from WikiTables from 2015 (called
Synthetic in Table~\ref{tab:benchmark}).
For each table, we randomly select some rows and insert them into a new synthetic table in random order.
We used this method to generate 1,494,290 new tables which we split into three corpora of different sizes: 500K, 1M, and all 1,494,290 tables.
In each corpus, we include the original WikiTables, resulting in corpora of 738,038, 1,238,038, and 1,732,328 tables.

\subsection{Semantic Table Search Prefiltering}

\begin{figure*}[t]
    \includegraphics[width=\textwidth]{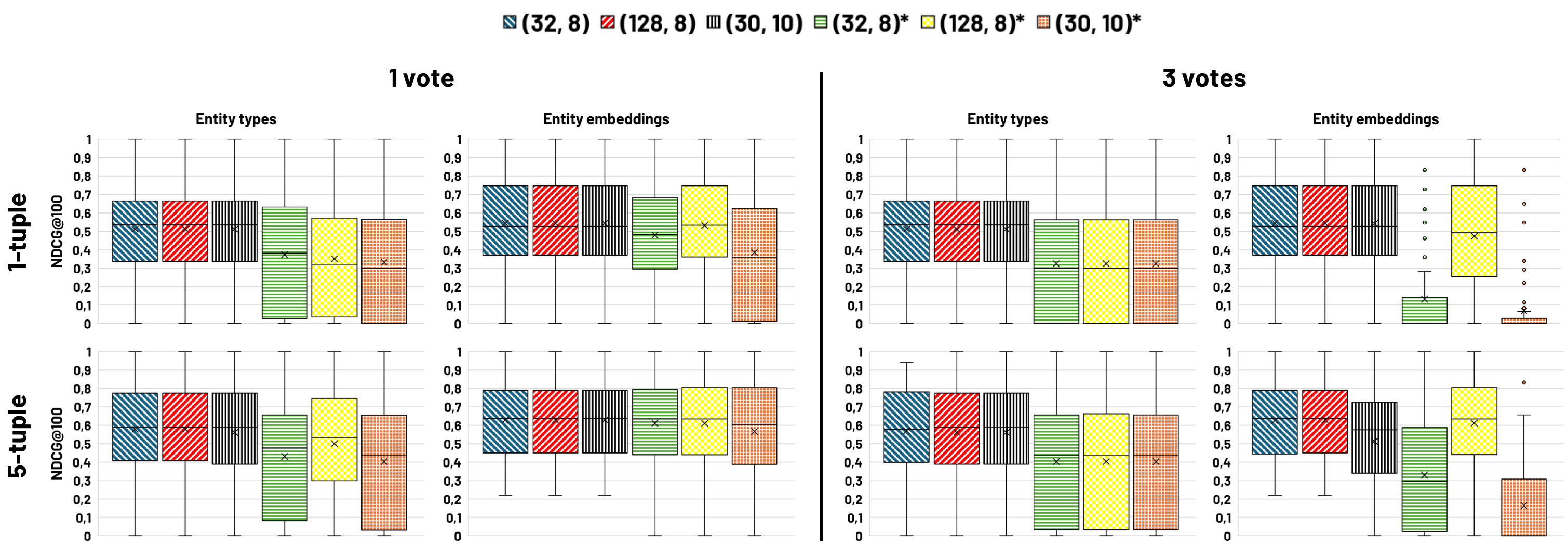}
    \caption{Semantic table search with LSH prefiltering using various LSH configurations: (number of permutation/projection vectors, band size) and * indicates that column aggregation is enabled}
    \label{fig:lsh_evaluation}
\end{figure*}

\begin{figure}[t]
    \includegraphics[width=\columnwidth]{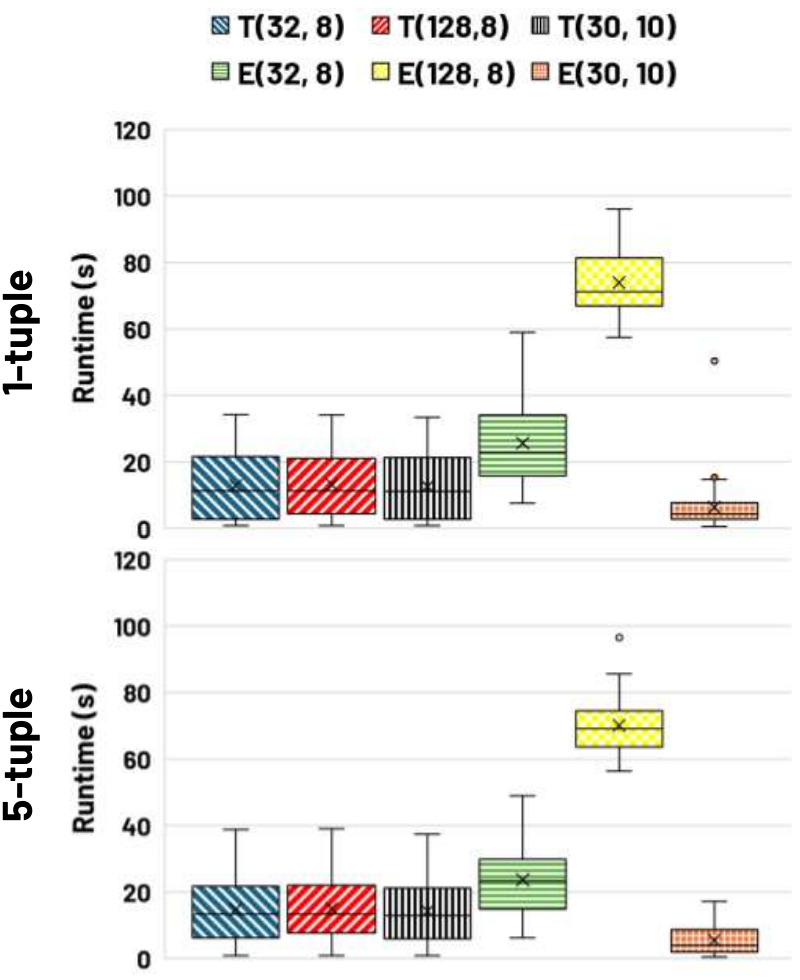}
    \caption{Runtime of semantic table search with LSH prefiltering using various LSH configurations: (number of permutation/projection vectors, band size), and \textit{T} means types and \textit{E} means embeddings}
    \label{fig:lsh_runtime}
\end{figure}

\begin{figure*}[t]
    \includegraphics[width=\textwidth]{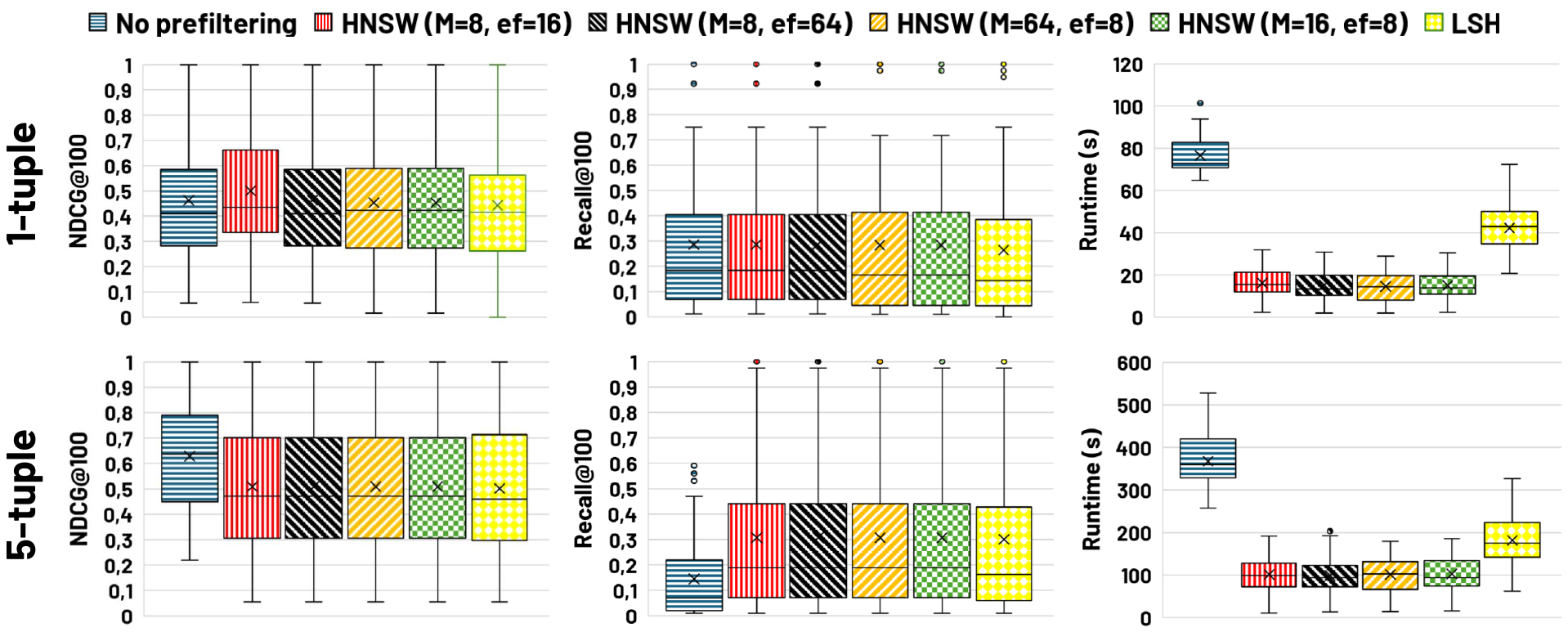}
    \caption{Semantic table search with HNSW prefiltering using various HNSW configurations compared to LSH and no prefiltering}
    \label{fig:hnsw_evaluation}
\end{figure*}

\begin{figure}[t]
    \centering
    \includegraphics[width=\linewidth]{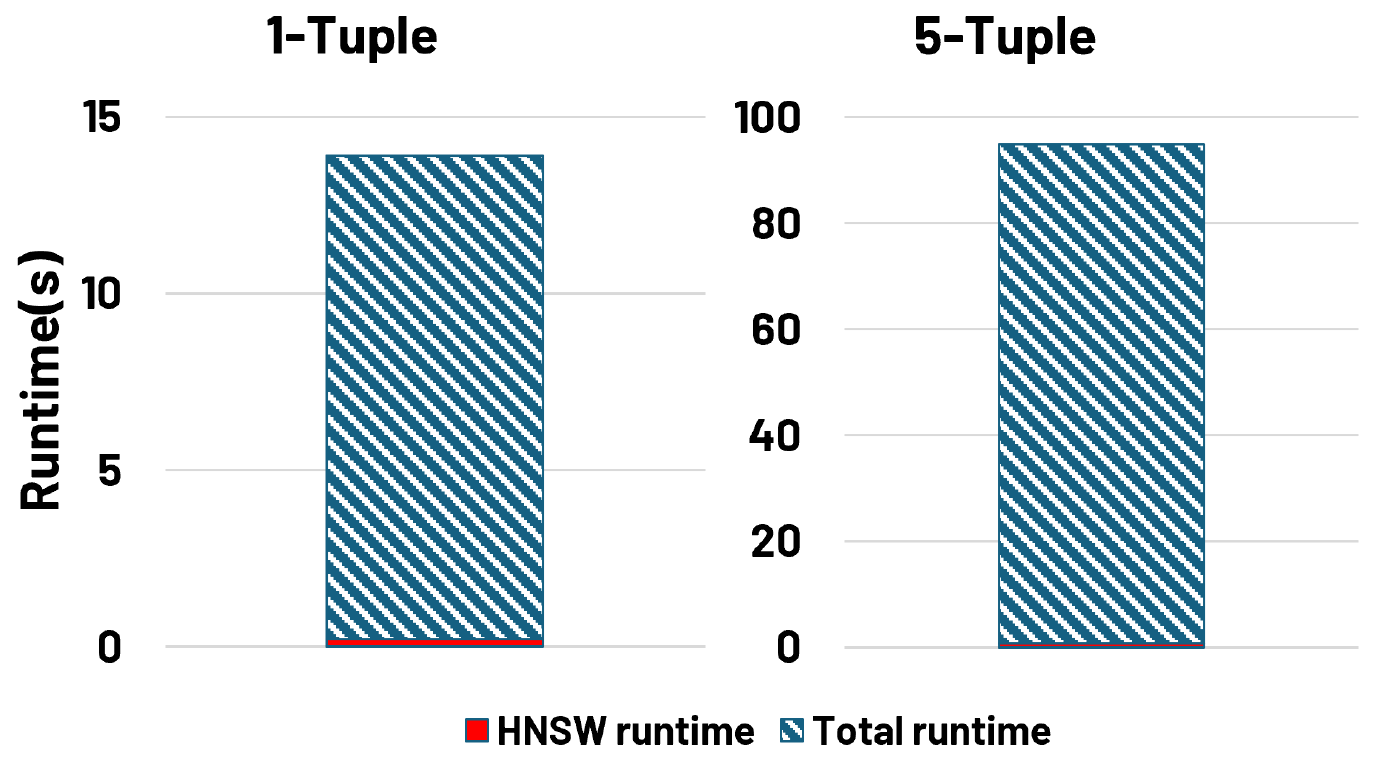}
    \caption{Runtime split between HNSW prefiltering and Thetis semantic table search}
    \label{fig:runtime_split}
\end{figure}

In the following, we first evaluate various LSH configurations.
Then, we evaluate various HNSW configurations and compare these to the best-performing LSH configuration and semantic search without prefiltering.

\noindent
\textbf{LSH Prefiltering.}
We compare the effect of 6 different LSH configurations when using entity types and embeddings.
This will inform the selection of our best configuration regarding the tradeoff between runtime reduction and output quality.
These configurations are denoted by $(X,Y)$, where $X$ is the number of permutation/projection vectors and $Y$ is the band size.
The configurations marked with an asterisk (*) apply column aggregation on tables during LSH indexing.
These LSH configurations have been selected after testing various configurations on a smaller subset of the corpus.
We evaluate the NDCG scores obtained with prefiltering with these configurations in Figure~\ref{fig:lsh_evaluation}.
All LSH configurations achieve equivalent NDCG scores as our semantic table search algorithm without prefiltering.
However, applying table column aggregation in LSH indexes achieves lower NDCG scores.
Because requiring 3 votes is more selective without degrading the ranking performance, we continue requiring 3 votes in the following experiments.

The runtime of \texttt{Thetis++} is correlated with the number of query tuples, as each query tuple is compared to all table rows.
Figure~\ref{fig:lsh_runtime} presents the runtimes of the LSH configurations using entity types and embeddings.
We see that the (30, 10)-configuration slightly outperforms the other two configurations, as this configuration has 4 times more buckets per bucket group than the other two configurations.
Therefore, each bucket contains fewer tables.
Furthermore, this configuration also has the lowest number of bucket groups, leading to a lower number of buckets to be merged into the final LSH result set.
Therefore, for the remainder of this section, we will only experiment with the (30, 10)-configuration, as this also has the best performance on other datasets.

We also evaluate a naive prefiltering technique using BM25 keyword search instead. 
On 1-tuple queries, NDCG decreased by 18.1\% compared to LSH using types and by 29.9\% compared to LSH using embeddings. 
On 5-tuple queries, NDCG decreased by 13.3\% and 13.4\%, respectively.
Hence, BM25 filters out many relevant tables and is not a valid prefiltering method.

\noindent
\textbf{HNSW Prefiltering.}
We evaluate various configurations of our HNSW index and again select the configuration that leads to the best semantic table search performance regarding output quality and runtime.
As previously described, HNSW has two main parameters: (1) $M$ is the number of new connections to a new node, and (2) $ef$ is the priority queue size during graph traversal (see Section~\ref{sec:approximate}).
Together, these parameters trade runtime and recall.
We select a set of 4 HNSW configurations and evaluate their impact on semantic table search output and ranking quality, as well as runtime.
We additionally compare to the best-performing LSH configuration and \texttt{Thetis++} without prefiltering.

Figure~\ref{fig:hnsw_evaluation} shows the NDCG, recall, and runtime of the HNSW configurations in comparison to semantic table search without prefiltering and when using LSH.
We see that the various HNSW configurations have a minor effect on the NDCG and recall.
Instead, all configurations retrieve a set of highly relevant tables overlapping with the retrieved tables without prefiltering on 1-tuple queries, indicated by the similar recall scores.
Note that one configuration improves the ranking performance of Thetis on 1-tuple queries, as the prefiltering with this configuration removed false positives.
This is especially the case for all HNSW configurations on 5-tuple queries, which is indicated by the higher recall.
However, the fewer relevant tables retrieved by Thetis without prefiltering have a higher ground truth ranking than those retrieved with HNSW prefiltering, and NDCG is sensitive to higher-ranked results; hence, the NDCG score without prefiltering is higher than with HNSW prefiltering.
The low recall of semantic table search without prefiltering specifically on 5-tuple queries is due to these queries being over-specialized.
Semantic table search can be sensitive towards single highly relevant entities, whereas this is not the case for the prefiltering techniques.
Therefore, in practice, the prefiltering is a pruning step for the case with large input queries.
HNSW prefiltering leads to a runtime improvement of up to 28$\times$ and on average 4.7$\times$ for 1-tuple queries with $M=8$ and $ef=16$, and 3.6$\times$ on average for 5-tuple queries.
HNSW also leads to a faster runtime than LSH, as its search space reduction is generally higher and the HNSW approximate nearest neighbor search is faster in general.
LSH improves runtime by up to 4.2$\times$.
\emph{Overall, these experiments show that the HNSW configurations do not harm the ranking and output quality of Thetis without prefiltering and even significantly improve runtime, including LSH prefiltering.}
Since the HNSW configuration with $M=8$ and $ef=16$ is slightly better than the other configuration, we will use this configuration for the remaining experiments for search space prefiltering.

Figure~\ref{fig:runtime_split} shows the total average runtime of Thetis semantic table search on 1- and 5-tuple queries as well as how much time was spent on HNSW search space prefiltering on average using the best HNSW configuration.
The average runtime spent prefiltering is 0.20s and 0.64s on 1- and 5-tuple queries, respectively.

\begin{figure}
    \centering
    \includegraphics[width=0.6\columnwidth]{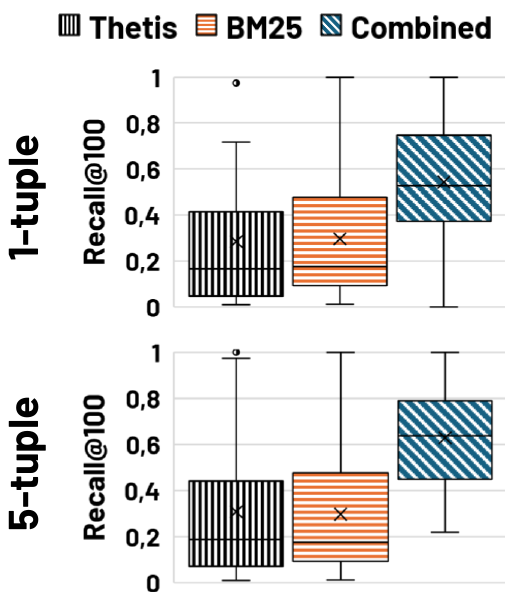}
    \caption{Recall when combining semantic table search and BM25}
    \label{fig:stse}
\end{figure}

\begin{figure*}[t]
    \centering
    \subfloat[NDCG of studied methods. TOPSIS($x$, $y$) denotes weights, where $x$ is the weight for Thetis and $y$ is the weight for BM25.]{%
        \includegraphics[width=\textwidth]{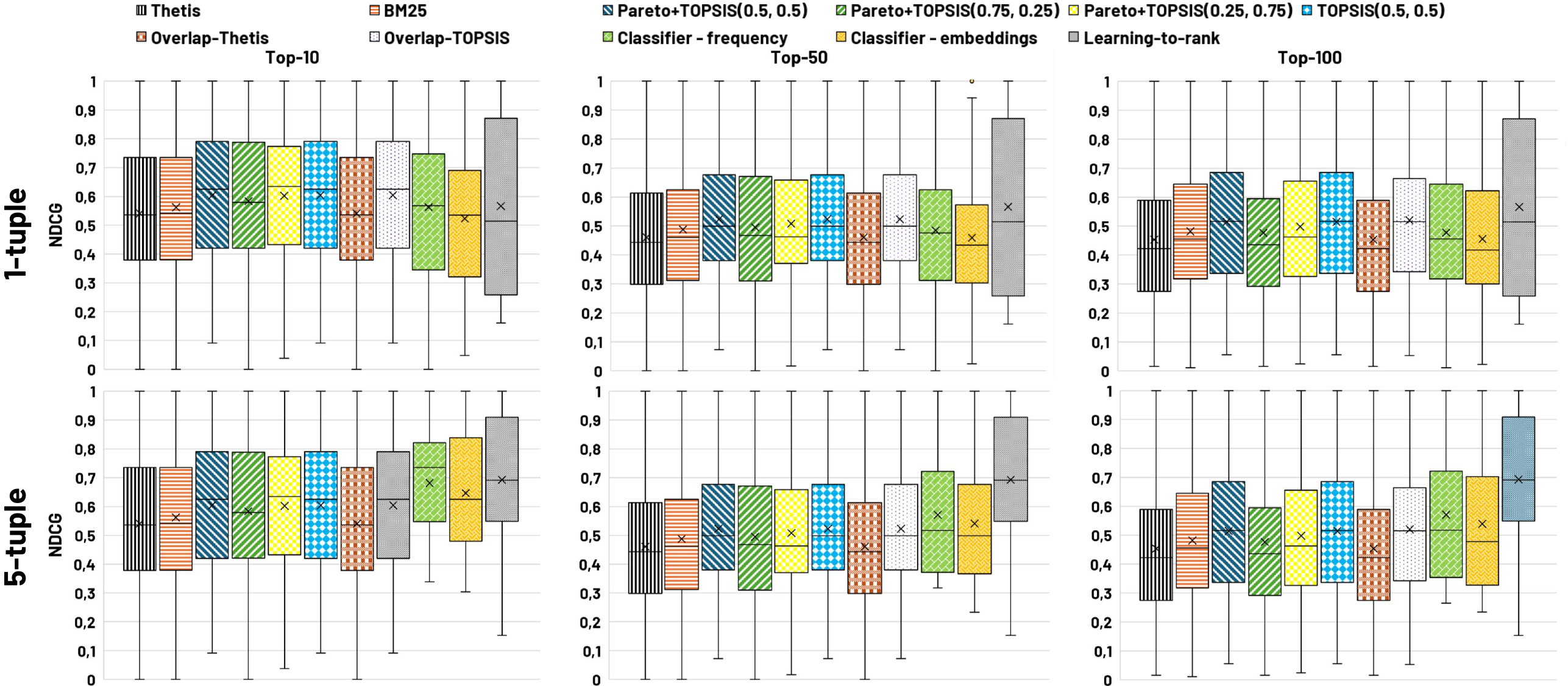}
        \label{fig:ndcg}
    }
    \\
    \subfloat[Recall of studied methods. TOPSIS($x$, $y$) denotes weights, where $x$ is the weight for Thetis and $y$ is the weight for BM25.]{%
        \includegraphics[width=\textwidth]{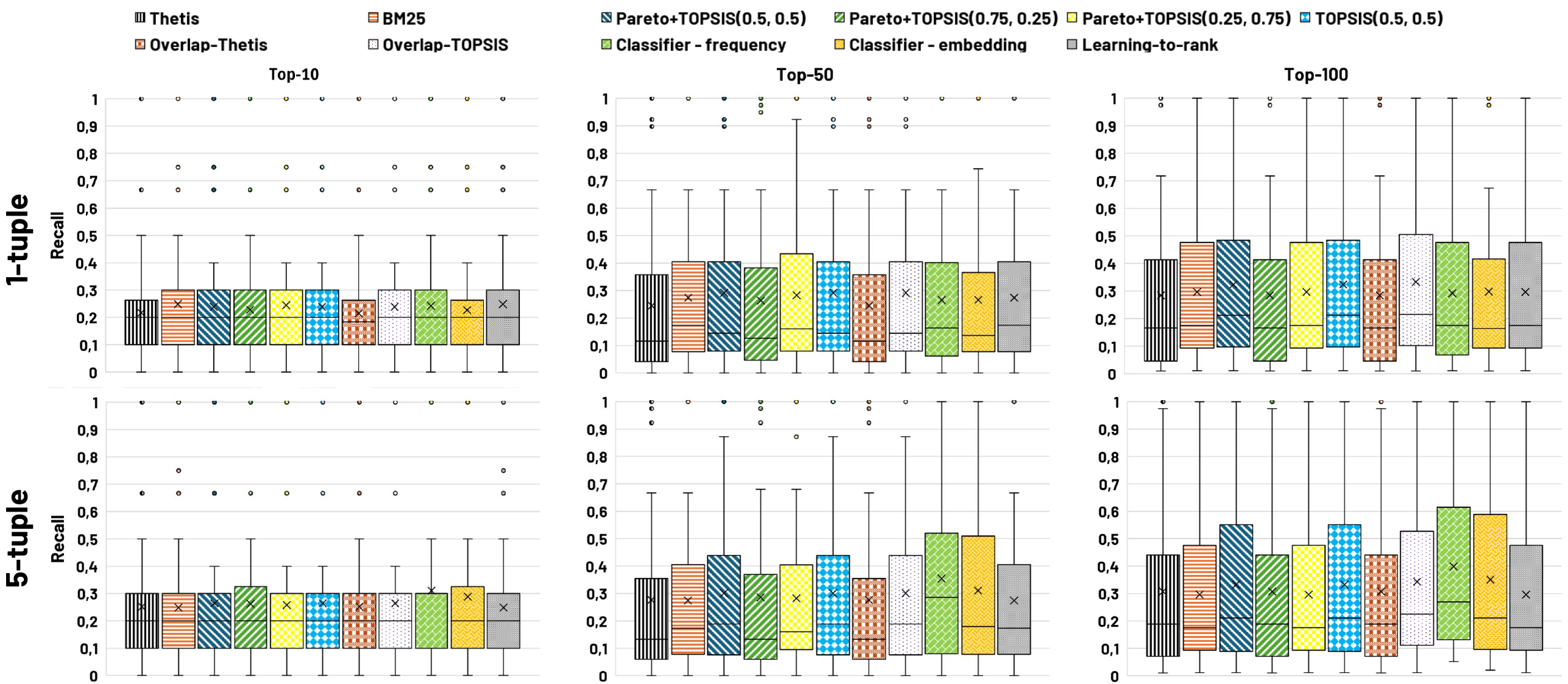}
        \label{fig:recall}
    }
    \caption{Ranking and output quality}
    \label{fig:evaluation}
\end{figure*}

\subsection{Semantic Table Search Quality}
We evaluate Thetis against BM25, Starmie, SANTOS, D$^3$L, and TURL.
We compare the NDCG scores (Figure~\ref{fig:evaluation}) as well as the improvement in recall separately and when using both techniques jointly.

Comparing NDCG and recall scores at top-10, 50 and 100 (Figure~\ref{fig:evaluation}), Thetis shows a similar ranking quality, i.e., NDCG and recall, to BM25 text queries on both 1- and 5-tuple queries.
This is due to BM25 effectively finding the ground truth relevant tables containing exact matches with the text queries, whereas Thetis finds a different set of ground truth relevant tables that do not necessarily contain exact matches.
Using a larger KG, such as WikiData, would expectedly result in better performance of Thetis, as WikiData is more detailed and descriptive than DBpedia for many entities.
Note that the performance of Thetis depends on the quality of the embeddings.
Similarly, Thetis using types depends on sufficiently fine-grained sets of entity types.
Moreover, Thetis using embeddings works well when the taxonomy is less detailed; however, the embeddings are at times not able to distinguish entities of the same type or domain, e.g., two countries or a country and its capital.
Despite the similar performance in NDCG and recall between Thetis and BM25, \emph{the two approaches find remarkably different subsets of relevant tables} (see Tables~\ref{tab:overlap_top_100_1_tuple}, \ref{tab:overlap_top_100_5_tuple}, ~\ref{tab:overlap_top_10_1_tuple}, and~\ref{tab:overlap_top_10_5_tuple}). 
Similar aggregate effectiveness scores mask the fact that the methods retrieve different relevant tables. 
This distinction is important because NDCG and recall alone do not reveal whether two methods retrieve the same or different relevant results.
Tables~\ref{tab:overlap_top_100_1_tuple}, \ref{tab:overlap_top_100_5_tuple}, ~\ref{tab:overlap_top_10_1_tuple}, and~\ref{tab:overlap_top_10_5_tuple} show the average overlap for top-10 and top-100.
These overlaps show that our semantic table search algorithm finds largely distinct set of tables from BM25.
We therefore study the benefit of a Semantic Data Lake when combining Semantic Table Search and BM25.
First, we evaluate a simple combination of our semantic table search algorithm and BM25: we extracted the top 50\% from each method, merged the two result sets, and measured recall at top-100.
We provide the results in Figure~\ref{fig:stse}, which shows a significantly improved recall.
Specifically, for top-100 on 1-tuple queries, the median recall is improved by 9.1\% to BM25.
On 5-tuple queries, recall is improved by 156.6\% to BM25.
Hence, complementing tables found by exact matching with tables found by semantic relevance combines the best of both worlds.

Due to the poor performances of join and union search, as well as representation learning, we leave out these approaches from our plots.
Specifically, union search with Starmie achieves an average NDCG of just 0.10 and recall of 0.03 on 1-tuple queries and 0.13 NDCG and 0.04 recall on 5-tuple queries, as relevant tables are often not unionable.
The performance of SANTOS and D$^3$L as representative union and join search approaches, respectively, is even worse with NDCG scores $\sim$1000x lower than Thetis.
Specifically, SANTOS achieves an average NDCG score of 0.0001 for both 1- and 5-tuple queries, and D$^3$L achieves 0.00006 and 0 for 1- and 5-tuple queries, respectively.
Their poor performances are due to these methods not being designed to take into account topical relevance, as described in Section~\ref{sec:soa_ts}.
Therefore, these methods are not able to rank tables according to semantic relevance, and thereby they fail to properly rank relevant tables, as experiments show.
However, Starmie obtains an improved performance, as it is capable of capturing rich contextual semantic information from tables using trained column encoders.
Similarly, TURL achieves average NDCG scores of 0.004 and 0.005 for the same queries, respectively.
However, TURL's performance can reach 0.488 using entire source tables.
This reflects that TURL is not designed for semantic search, but rather for table understanding.

\subsection{Combining Keyword and Semantic Search}
We evaluate \texttt{Thetis++} (\texttt{P+T}) using 3 different weights when combining Thetis (A) semantic table search and BM25 (B) keyword search.
We furthermore study the various alternative techniques and compare their performance to \texttt{Thetis++}, Thetis, and BM25.
This evaluation is carried out using NDCG and recall, and the results are in Figure~\ref{fig:evaluation}.

The ranking performance in Figure~\ref{fig:ndcg} shows a similar performance by every evaluated technique.
Compared with Thetis and BM25, almost all multi-criteria methods slightly improve NDCG, except \texttt{Overlap-Thetis} (\texttt{O+A}).
Therefore, Thetis is not as effective in providing the correct ranking scores in comparison to TOPSIS, which computes a combined ranking score using both Thetis and BM25 relevance scores.
Furthermore, we see that the multi-criteria ranking techniques improve ranking with Thetis or BM25 to a smaller degree as we increase the result set size, i.e., larger result sets lead to a more similar ranking performance across all methods, including Thetis and BM25.
\emph{This shows that the multi-criteria ranking is more effective at providing correct rankings of the highest ranking tables, whereas the lower ranked tables do not provide clear enough relevance signals from each search method to provide accurate, individual relevance scores.}

Considering the ranking performances of \texttt{Pareto+TOPSIS} using different weights, we see that the best performance is achieved when using equal weights for both Thetis and BM25.
\emph{This highlights that both methods contribute to a high-quality ranking equally and that both methods should be considered when combining their rankings.}
\texttt{Learning-to-rank} (L) achieves a less stable NDCG with an average and median NDCG score similar to the other approach when provided with small queries.
However, \emph{the performance improves as we increase query sizes to include more information}.
Furthermore, \emph{many queries achieve a higher NDCG than any other evaluated approach, highlighting its ability to better rank keyword search-retrieved tables than BM25}.
Its performance is stable across all top-\textit{K} result set sizes.
However, its overall poor and unstable performance points towards experimenting with Reciprocal Rank Fusion (RFF)~\cite{fadha2026comparing, rff, Bruch_2023} in the future as an alternative method for multi-criteria ranking, as RFF has shown potential in previous evaluations.

Recall for small result sets for every technique is near identical, and we must consider larger result sets before the performances begin to differ.
For larger result sets, we see a similar trend, where multi-criteria ranking techniques obtain only a slightly improved performance.
Therefore, \emph{these techniques struggle to replace false positives in one search method with true positives from another method}.
Instead, it commonly occurs in all of these techniques that a true positive in one method is replaced with a true positive from another method (e.g., a true positive retrieved with Thetis is replaced with a true positive retrieved with BM25).
Note that the initial set of tables to be ranked by \texttt{Learning-to-Rank} is constructed from BM25 keyword search, and hence, the recall remains the same.

\begin{table*}[h]
\caption{Fractions of overlapping tables also included in the ground truth for top-100 1-tuple query results: A = Thetis, B = BM25, P = Pareto, T = TOPSIS - TOPSIS($x$) is given weight $0<x\leq1$ for Thetis search, and BM25 is weighted $1 - x$ O = Overlap, C = Classifier, F = frequency, E = embeddings, and L = Learning-to-rank}
\label{tab:overlap_top_100_1_tuple}
\adjustbox{width=\textwidth}{
\begin{tabular}{|
>{\columncolor[HTML]{C0C0C0}}l |r|r|r|r|r|r|r|r|r|r|r|}
\hline
\textbf{}                          & \cellcolor[HTML]{C0C0C0}\textbf{A} & \cellcolor[HTML]{C0C0C0}\textbf{B} & \cellcolor[HTML]{C0C0C0}\textbf{P/T(0.5)} & \cellcolor[HTML]{C0C0C0}\textbf{P/T(0.75)} & \cellcolor[HTML]{C0C0C0}\textbf{P/T(0.25)} & \cellcolor[HTML]{C0C0C0}\textbf{T(0.5)} & \cellcolor[HTML]{C0C0C0}\textbf{O/A} & \cellcolor[HTML]{C0C0C0}\textbf{O/T} & \cellcolor[HTML]{C0C0C0}\textbf{C/F} & \cellcolor[HTML]{C0C0C0}\textbf{C/E} & \cellcolor[HTML]{C0C0C0}\textbf{L} \\ \hline
\textbf{A}                    & 1.00                                    & 0.07                                  & \underline{0.11}                                                     & \textbf{0.13}                                                       & 0.07                                                       & \underline{0.11}                                              & \textbf{0.13}                                            & \underline{0.11}                                            & 0.08                                                        & 0.09                                                          & 0.07                                                  \\ \hline
\textbf{B}                      & 0.07                                    & 1.00                                  & 0.11                                                     & 0.07                                                       & \textbf{0.14}                                                       & 0.11                                              & 0.07                                            & \underline{0.12}                                            & 0.10                                                        & 0.11                                                          & 1.00                                                  \\ \hline
\textbf{P/T(0.5)}   & 0.11                                    & 0.11                                  & 1.00                                                     & 0.11                                                       & 0.11                                                       & \textbf{0.15}                                              & 0.11                                            & \underline{0.14}                                            & 0.10                                                        & 0.11                                                          & 0.11                                                  \\ \hline
\textbf{P/T(0.75)} & \textbf{0.13}                                    & 0.07                                  & \underline{0.11}                                                     & 1.00                                                       & 0.07                                                       & \underline{0.11}                                              & \textbf{0.13}                                            & \underline{0.11}                                            & 0.08                                                        & 0.09                                                          & 0.07                                                  \\ \hline
\textbf{P/T(0.25)} & 0.07                                    & \textbf{0.14}                                  & 0.11                                                     & 0.07                                                       & 1.00                                                       & 0.11                                              & 0.07                                            & \underline{0.12}                                            & 0.10                                                        & 0.11                                                          & \textbf{0.14}                                                  \\ \hline
\textbf{T(0.5)}          & 0.11                                    & 0.11                                  & \textbf{0.15}                                                     & 0.11                                                       & 0.11                                                       & 1.00                                              & 0.11                                            & \underline{0.14}                                            & 0.10                                                        & 0.11                                                          & 0.11                                                  \\ \hline
\textbf{O/A}            & \textbf{0.13}                                    & 0.07                                  & \underline{0.11}                                                     & \textbf{0.13}                                                       & 0.07                                                       & \underline{0.11}                                              & 1.00                                            & \underline{0.11}                                            & 0.08                                                        & 0.09                                                          & 0.07                                                  \\ \hline
\textbf{O/T}            & 0.11                                    & \underline{0.12}                                  & \textbf{0.14}                                                     & 0.11                                                       & \underline{0.12}                                                       & \textbf{0.14}                                              & 0.11                                            & 1.00                                            & 0.11                                                        & 0.11                                                          & \underline{0.12}                                                  \\ \hline
\textbf{C/F}    &                                        0.08 &                                      \underline{0.10} &                                                         \underline{0.10} &                                                           0.08 &                                                           \underline{0.10} &                                                  \underline{0.10} &                                                0.08 &                                                \textbf{0.11} &                                                        1.00 &                                                         \underline{0.10} & \underline{0.10}                                                 \\ \hline
\textbf{C/E}   & 0.09                                        & \textbf{0.11}                                      & \textbf{0.11}                                                         & 0.09                                                           & \textbf{0.11}                                                           & \textbf{0.11}                                                  & 0.09                                                & \textbf{0.11}                                                & \underline{0.10}                                                        & 1.00                                                          & \textbf{0.11}                                                  \\ \hline
\textbf{L}          & 0.07                                    & 1.00                                  & 0.11                                                     & 0.07                                                       & \textbf{0.14}                                                       & 0.11                                              & 0.07                                            & \underline{0.12}                                            & 0.10                                                        & 0.11                                                          & 1.00                                                  \\ \hline
\end{tabular}}
\end{table*}

\begin{table*}[ht]
\caption{Fractions of overlapping tables also included in the ground truth for top-100 5-tuple query results: A = Thetis, B = BM25, P = Pareto, T = TOPSIS - TOPSIS($x$) is given weight $0<x\leq1$ for Thetis search, and BM25 is weighted $1 - x$ O = Overlap, C = Classifier, F = frequency, E = embeddings, and L = Learning-to-rank}
\label{tab:overlap_top_100_5_tuple}
\adjustbox{width=\textwidth}{
\begin{tabular}{|
>{\columncolor[HTML]{C0C0C0}}l |r|r|r|r|r|r|r|r|r|r|r|}
\hline
\textbf{}                          & \cellcolor[HTML]{C0C0C0}\textbf{A} & \cellcolor[HTML]{C0C0C0}\textbf{B} & \cellcolor[HTML]{C0C0C0}\textbf{P/T(0.5)} & \cellcolor[HTML]{C0C0C0}\textbf{P/T(0.75)} & \cellcolor[HTML]{C0C0C0}\textbf{P/T(0.25)} & \cellcolor[HTML]{C0C0C0}\textbf{T(0.5)} & \cellcolor[HTML]{C0C0C0}\textbf{O/A} & \cellcolor[HTML]{C0C0C0}\textbf{O/T} & \cellcolor[HTML]{C0C0C0}\textbf{C/F} & \cellcolor[HTML]{C0C0C0}\textbf{C/E} & \cellcolor[HTML]{C0C0C0}\textbf{L} \\ \hline
\textbf{A}                    & 1.00                                    & 0.08                                  & 0.12                                                     & \textbf{0.15}                                                       & 0.08                                                       & 0.12                                              & \textbf{0.15}                                            & \underline{0.13}                                            & 0.09                                                        & 0.11                                                          & 0.08                                                  \\ \hline
\textbf{B}                      & 0.08                                    & 1.00                                  & \underline{0.12}                                                     & 0.08                                                       & \textbf{0.14}                                                       & \underline{0.12}                                              & 0.08                                            & \underline{0.12}                                            & \underline{0.12}                                                        & 0.09                                                          & 1.00                                                  \\ \hline
\textbf{P/T(0.5)}   & 0.12                                    & 0.12                                  & 1.00                                                     & 0.12                                                       & 0.12                                                       & \textbf{0.16}                                              & 0.12                                            & \underline{0.15}                                            & 0.11                                                        & 0.11                                                          & 0.12                                                  \\ \hline
\textbf{P/T(0.75)} & \textbf{0.15}                                    & 0.08                                  & 0.12                                                     & 1.00                                                       & 0.08                                                       & 0.12                                              & \textbf{0.15}                                            & \underline{0.13}                                            & 0.09                                                        & 0.11                                                          & 0.08                                                  \\ \hline
\textbf{P/T(0.25)} & 0.08                                    & \textbf{0.14}                                  & \underline{0.12}                                                     & 0.08                                                       & 1.00                                                       & \underline{0.12}                                              & 0.08                                            & \underline{0.12}                                            & \underline{0.12}                                                        & 0.09                                                          & \textbf{0.14}                                                  \\ \hline
\textbf{T(0.5)}          & 0.12                                    & 0.12                                  & \textbf{0.16}                                                     & 0.12                                                       & 0.12                                                       & 1.00                                              & 0.12                                            & \underline{0.15}                                            & 0.11                                                        & 0.11                                                          & 0.12                                                  \\ \hline
\textbf{O/A}            & \textbf{0.15}                                    & 0.08                                  & 0.12                                                     & \textbf{0.15}                                                       & 0.08                                                       & 0.12                                              & 1.00                                            & \underline{0.13}                                            & 0.09                                                        & 0.11                                                          & 0.08                                                  \\ \hline
\textbf{O/T}            & \underline{0.13}                                    & 0.12                                  & \textbf{0.15}                                                     & \underline{0.13}                                                       & 0.12                                                       & \textbf{0.15}                                              & \underline{0.13}                                            & 1.00                                            & 0.12                                                        & 0.11                                                          & 0.12                                                  \\ \hline
\textbf{C/F}    &                                        0.09 &                                      \textbf{0.12} &                                                         \underline{0.11} &                                                           0.09 &                                                           \textbf{0.12} &                                                  \underline{0.11} &                                                0.09 &                                                \textbf{0.12} &                                                        1.00 &                                                         \underline{0.11} & \textbf{0.12}                                                  \\ \hline
\textbf{C/E}   & \textbf{0.11}                                        & \underline{0.09}                                      & \textbf{0.11}                                                         & \textbf{0.11}                                                           & \underline{0.09}                                                           & \textbf{0.11}                                                  & \textbf{0.11}                                                & \textbf{0.11}                                                & \textbf{0.11}                                                        & 1.00                                                          & \underline{0.09}                                                  \\ \hline
\textbf{L}          & 0.08                                    & 1.00                                  & \underline{0.12}                                                     & 0.08                                                       & \textbf{0.14}                                                       & \underline{0.12}                                              & 0.08                                            & \underline{0.12}                                            & \underline{0.12}                                                        & 0.09                                                          & 1.00                                                  \\ \hline
\end{tabular}}
\end{table*}

\begin{table*}[ht]
\caption{Fractions of overlapping tables also included in the ground truth for top-10 1-tuple query results: A = Thetis, B = BM25, P = Pareto, T = TOPSIS - TOPSIS($x$) is given weight $0<x\leq1$ for Thetis search, and BM25 is weighted $1 - x$ O = Overlap, C = Classifier, F = frequency, E = embeddings, and L = Learning-to-rank}
\label{tab:overlap_top_10_1_tuple}
\adjustbox{width=\textwidth}{
\begin{tabular}{|
>{\columncolor[HTML]{C0C0C0}}l |r|r|r|r|r|r|r|r|r|r|r|}
\hline
\textbf{}                          & \cellcolor[HTML]{C0C0C0}\textbf{A} & \cellcolor[HTML]{C0C0C0}\textbf{B} & \cellcolor[HTML]{C0C0C0}\textbf{P/T(0.5)} & \cellcolor[HTML]{C0C0C0}\textbf{P/T(0.75)} & \cellcolor[HTML]{C0C0C0}\textbf{P/T(0.25)} & \cellcolor[HTML]{C0C0C0}\textbf{T(0.5)} & \cellcolor[HTML]{C0C0C0}\textbf{O/A} & \cellcolor[HTML]{C0C0C0}\textbf{O/T} & \cellcolor[HTML]{C0C0C0}\textbf{C/F} & \cellcolor[HTML]{C0C0C0}\textbf{C/E} & \cellcolor[HTML]{C0C0C0}\textbf{L} \\ \hline
\textbf{A}                    & 1.00                                    & 0.19                                  & 0.27                                                     & \underline{0.30}                                                       & 0.25                                                       & 0.27                                              & \textbf{0.38}                                            & 0.27                                            & 0.24                                                        & 0.23                                                          & 0.19                                                  \\ \hline
\textbf{B}                      & 0.19                                    & 1.00                                  & \underline{0.30}                                                     & 0.24                                                       & \textbf{0.32}                                                       & \underline{0.30}                                              & 0.18                                            & \underline{0.30}                                            & 0.25                                                        & 0.29                                                          & 1.00                                                  \\ \hline
\textbf{P/T(0.5)}   & 0.27                                    & 0.30                                  & 1.00                                                     & 0.39                                                       & \underline{0.42}                                                       & \textbf{0.45}                                              & 0.27                                            & \textbf{0.45}                                            & 0.28                                                        & 0.26                                                          & 0.30                                                  \\ \hline
\textbf{P/T(0.75)} & 0.30                                    & 0.24                                  & \textbf{0.39}                                                     & 1.00                                                       & \underline{0.36}                                                       & \textbf{0.39}                                              & 0.29                                            & \textbf{0.39}                                            & 0.27                                                        & 0.25                                                          & 0.24                                                  \\ \hline
\textbf{P/T(0.25)} & 0.25                                    & 0.32                                  & \textbf{0.42}                                                     & \underline{0.36}                                                       & 1.00                                                       & \textbf{0.42}                                              & 0.25                                            & \textbf{0.42}                                            & 0.29                                                        & 0.28                                                          & 0.32                                                  \\ \hline
\textbf{T(0.5)}          & 0.27                                    & 0.30                                  & \textbf{0.45}                                                     & 0.39                                                       & \underline{0.42}                                                       & 1.00                                              & 0.27                                            & \textbf{0.45}                                            & 0.28                                                        & 0.26                                                          & 0.30                                                  \\ \hline
\textbf{O/A}            & \textbf{0.38}                                    & 0.18                                  & 0.27                                                     & \underline{0.29}                                                       & 0.25                                                       & 0.27                                              & 1.00                                            & 0.27                                            & 0.24                                                        & 0.22                                                          & 0.18                                                  \\ \hline
\textbf{O/T}            & 0.27                                    & 0.30                                  & \textbf{0.45}                                                     & 0.39                                                       & \underline{0.42}                                                       & \textbf{0.45}                                              & 0.27                                            & 1.00                                            & 0.28                                                        & 0.26                                                          & 0.30                                                  \\ \hline
\textbf{C/F}    & 0.24                                        & 0.25                                      & \underline{0.28}                                                         & 0.27                                                           & \textbf{0.29}                                                           & \underline{0.28}                                                  & 0.24                                                & \underline{0.28}                                                & 1.00                                                        & \underline{0.28}                                                          & 0.25                                                  \\ \hline
\textbf{C/E}   & 0.23                                        & \textbf{0.29}                                      & 0.26                                                         & 0.25                                                           & \underline{0.28}                                                           & 0.26                                                  & 0.22                                                & 0.26                                                & \underline{0.28}                                                        & 1.00                                                          & \textbf{0.29}                                                  \\ \hline
\textbf{L}          & 0.19                                    & 1.00                                  & \underline{0.30}                                                     & 0.24                                                       & \textbf{0.32}                                                       & \underline{0.30}                                              & 0.18                                            & \underline{0.30}                                            & 0.25                                                        & 0.29                                                          & 1.00                                                  \\ \hline
\end{tabular}}
\end{table*}

\begin{table*}[ht]
\caption{Fractions of overlapping tables also included in the ground truth for top-10 5-tuple query results: A = Thetis, B = BM25, P = Pareto, T = TOPSIS - TOPSIS($x$) is given weight $0<x\leq1$ for Thetis search, and BM25 is weighted $1 - x$ O = Overlap, C = Classifier, F = frequency, E = embeddings, and L = Learning-to-rank}
\label{tab:overlap_top_10_5_tuple}
\adjustbox{width=\textwidth}{
\begin{tabular}{|
>{\columncolor[HTML]{C0C0C0}}l |r|r|r|r|r|r|r|r|r|r|r|}
\hline
\textbf{}                          & \cellcolor[HTML]{C0C0C0}\textbf{A} & \cellcolor[HTML]{C0C0C0}\textbf{B} & \cellcolor[HTML]{C0C0C0}\textbf{P/T(0.5)} & \cellcolor[HTML]{C0C0C0}\textbf{P/T(0.75)} & \cellcolor[HTML]{C0C0C0}\textbf{P/T(0.25)} & \cellcolor[HTML]{C0C0C0}\textbf{T(0.5)} & \cellcolor[HTML]{C0C0C0}\textbf{O/A} & \cellcolor[HTML]{C0C0C0}\textbf{O/T} & \cellcolor[HTML]{C0C0C0}\textbf{C/F} & \cellcolor[HTML]{C0C0C0}\textbf{C/E} & \cellcolor[HTML]{C0C0C0}\textbf{L} \\ \hline
\textbf{A}                    & 1.00                                    & 0.20                                  & 0.29                                                     & \underline{0.34}                                                       & 0.26                                                       & 0.29                                              & \textbf{0.43}                                            & 0.29                                            &                                                        0.32 &                                                          \underline{0.34} & 0.20                                                   \\ \hline
\textbf{B}                      & 0.20                                    & 1.00                                  & \underline{0.32}                                                     & 0.27                                                       & \textbf{0.33}                                                       & \underline{0.32}                                              & 0.20                                            & \underline{0.32}                                            & 0.31                                                        & 0.27                                                          & 1.00                                                  \\ \hline
\textbf{P/T(0.5)}   & 0.29                                    & 0.32                                  & 1.00                                                     & 0.40                                                       & \underline{0.43}                                                       & \textbf{0.47}                                              & 0.29                                            & \textbf{0.47}                                            & 0.36                                                        & 0.34                                                          & 0.32                                                  \\ \hline
\textbf{P/T(0.75)} & 0.34                                    & 0.27                                  & \underline{0.40}                                                     & 1.00                                                       & 0.37                                                       & \textbf{0.41}                                              & 0.34                                            & \textbf{0.41}                                            & 0.35                                                        & 0.35                                                          & 0.27                                                  \\ \hline
\textbf{P/T(0.25)} & 0.26                                    & 0.33                                  & \textbf{0.43}                                                     & \underline{0.37}                                                       & 1.00                                                       & \textbf{0.43}                                              & 0.26                                            & \textbf{0.43}                                            & 0.35                                                        & 0.33                                                          & 0.33                                                  \\ \hline
\textbf{T(0.5)}          & 0.29                                    & 0.32                                  & \textbf{0.47}                                                     & 0.41                                                       & \underline{0.43}                                                       & 1.00                                              & 0.29                                            & \textbf{0.47}                                            & 0.36                                                        & 0.34                                                          & 0.32                                                  \\ \hline
\textbf{O/A}            & \textbf{0.43}                                    & 0.20                                  & 0.29                                                     & \underline{0.34}                                                       & 0.26                                                       & 0.29                                              & 1.00                                            & 0.29                                            & 0.32                                                        & \underline{0.34}                                                          & 0.20                                                  \\ \hline
\textbf{O/T}            & 0.29                                    & 0.32                                  & \textbf{0.47}                                                     & 0.41                                                       & \underline{0.43}                                                       & \textbf{0.47}                                              & 0.29                                            & 1.00                                            & 0.36                                                        & 0.34                                                          & 0.32                                                  \\ \hline
\textbf{C/F}    &                                        0.32 &                                      0.31 &                                                        
\underline{0.36} &                                                           0.35 &                                                           0.35 &                                                  \underline{0.36} &                                                0.23 &                                                \underline{0.36} &                                                        1.00 & \textbf{0.43}                                                          & 0.31                                                  \\ \hline
\textbf{C/E}   & 0.34                                        & 0.27                                      & 0.34                                                         & \underline{0.35}                                                           & 0.33                                                           & 0.34                                                  & 0.34                                                & 0.34                                                & \textbf{0.43}                                                        & 1.00                                                          & 0.27                                                  \\ \hline
\textbf{L}          & 0.20                                    & 1.00                                  & \underline{0.32}                                                     & 0.27                                                       & \textbf{0.33}                                                       & \underline{0.32}                                              & 0.20                                            & \underline{0.32}                                            & 0.31                                                        & 0.27                                                          & 1.00                                                  \\ \hline
\end{tabular}}
\end{table*}

Table~\ref{tab:overlap_top_100_1_tuple} shows the overlap fraction for top-100 result sets for each evaluated technique and ground truth on 1-tuple queries, and Table~\ref{tab:overlap_top_100_5_tuple} for 5-tuple queries.
In general, the overlaps are quite small, but we note that the overlaps expectedly increase between pairs of more similar techniques.
For example, Thetis has a higher overlap with \texttt{Pareto+TOPSIS}$(0.75, 0.25)$, as the latter weighs higher ranking scores provided by Thesis.
The overlap between \texttt{Pareto+TOPSIS} and TOPSIS is among the highest, but yet still unexpectedly low.
This shows the significant impact Pareto has on the final ranking.
In Tables~\ref{tab:overlap_top_10_1_tuple} and ~\ref{tab:overlap_top_10_5_tuple}, we show the same overlap for top-10 results on 1- and 5-tuple queries, respectively.
The overlaps here are significantly higher overall, which highlights that \emph{the most important tables are ranked highest}, whereas the various techniques struggle at selecting the right tables once both search methods do not provide high relevance scores.
This is also aligned with the plots in Figure~\ref{fig:ndcg}, where the performance for smaller result sets leads to higher NDCG scores.
Note that we compute the overlaps for top-10 using the results for top-100.
Therefore, after combining result sets for \texttt{Overlap-Thetis} and \texttt{Overlap-TOPSIS}, the tables retrieved with Thetis may be ranked lower and are therefore outside top-10, leading to a lower overlap fraction.
Even though \texttt{Overlap-Thetis} employs Thetis for ranking the combined result set, it might still rank its own retrieved tables lower than those retrieved with BM25.
This is because Thetis employs HNSW prefiltering, and BM25 can therefore introduce tables that were prefiltered and should be assigned a higher relevance score by Thetis.

\subsection{Runtime Evaluation}
We study the computation cost of the mapping function $\mu_{T,Q}$ (Section~\ref{subsec:algo}) and compare it to the total cost required to score a single table. 
We find that the total, average runtimes of scoring a WikiTables from 2015 table using 1- and 5-tuple queries are 2.2ms and 8.6ms, respectively.
Similarly, the average runtimes of scoring a GitTable on the same queries are 3.8ms and 16.6ms, respectively.
On WikiTables 2015, for 1-tuple quieres, 63.7\% and 58.6\% of runtime is spent computing $\mu_{T,Q}$  using types and embeddings, respectively.
On 5-tuple queries, these fractions are 74.5\% and 67.3\%.
Similarly, on GitTables, 68.0\% and 62.4\% are spent on this computation on 1-tuple queries, and 78.1\% and 75.7\% on 5-tuple queries.
Hence, the overall time spent for Algorithm~\ref{alg:generator} and the cost of $\mu_{T,Q}$ are limited, even when dealing with larger tables.

The methods based on classification as well as the learning-to-rank method require a training-phase of a machine learning model.
Training the classifier on 1K randomly sampled queries takes 2.6s on average.
The learning-to-rank method spends 3.8h training its model on the same set of queries.

\begin{figure}
    \centering
    \includegraphics[width=\columnwidth]{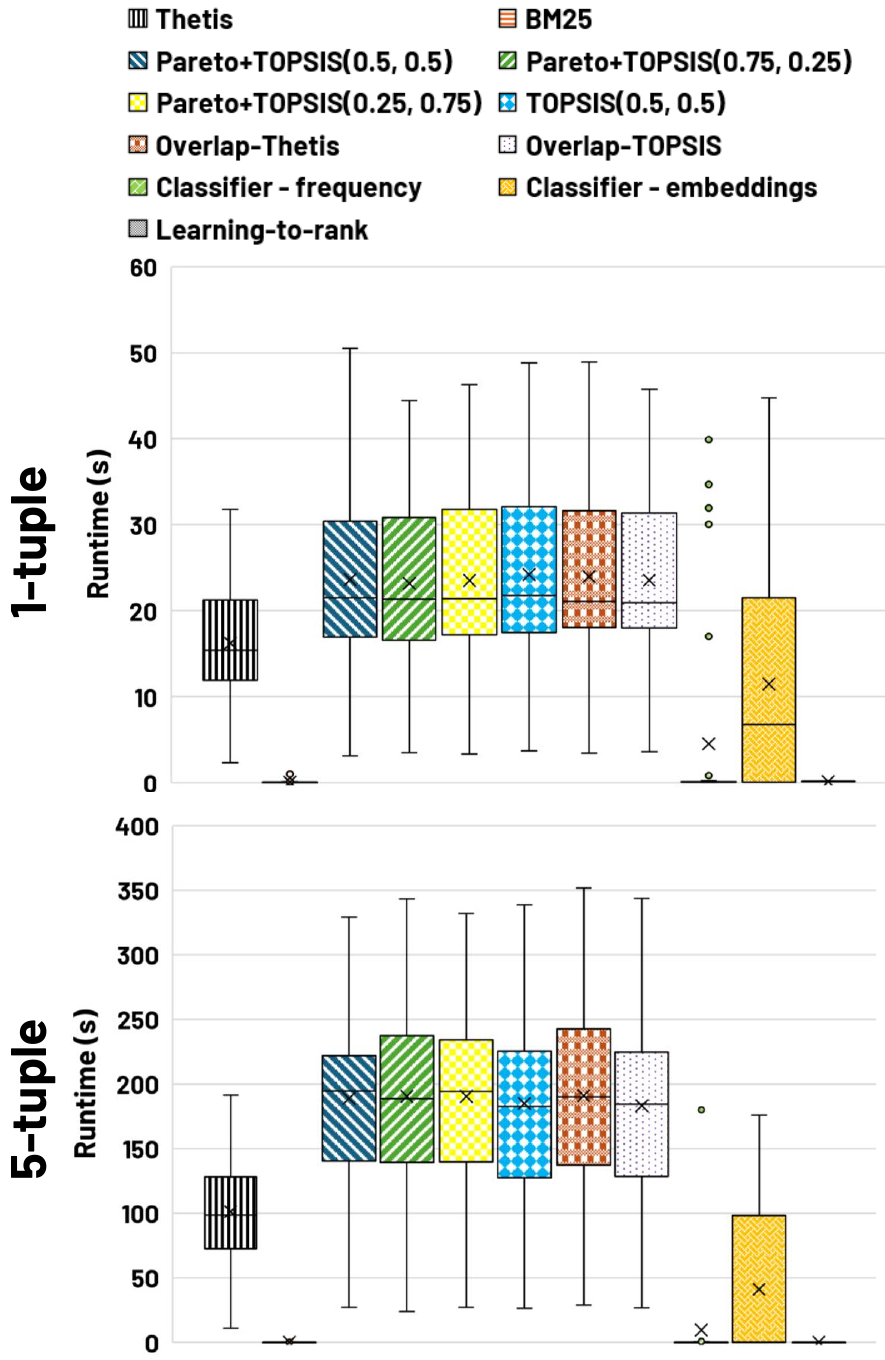}
    \caption{Runtimes of data lake table search methods}
    \label{fig:runtime}
\end{figure}

Figure~\ref{fig:runtime} plots the runtimes of the multi-criteria ranking methods as well as Thetis and BM25.
Thetis employs prefiltering using HNSW, and this prefiltering is also employed in all the other multi-criteria ranking methods that employ Thetis.
BM25 is expectedly much faster than Thetis, as it relies on inverted indexes.
Every method that uses Pareto or TOPSIS relies on Thetis, resulting in runtimes at least as slow as Thetis.
Due to additional processing of the ranked tables, the runtimes of these methods are slightly higher than those of Thetis itself.
The classifier techniques predict whether to search with Thetis or BM25.
Given that \texttt{Classifier - frequency} obtains a runtime nearly the same as BM25, it implies that the majority of its predictions decided on using BM25 for ranking.
On the contrary, \texttt{Classifier - embeddings} obtains a runtime only slightly better than Thetis, which implies it mostly predicted to use Thetis for ranking.
Learning-to-rank achieved the best runtime, but at the expense of a runtime-costly training phase.

\subsection{Experiments on Other Datasets}
The following experiments are performed with LSH prefiltering using entity types and embeddings~\cite{christensen2025fantastic}.

\noindent
\textbf{Synthetic Dataset.}
Using 0.7M tables, the runtimes on 1-tuple queries are on average 20.7s using types and 56.2s using embeddings.
These runtimes increase to 27.4s and 71.8s on 1.2M tables and 29.9s and 85.1s on 1.7M tables.
The same linear runtime increase is observed using 5-tuple queries:
54.4s and 282.8s on 0.7M tables, 73.4s and 343.4s on 1.2M tables, and 77.3s and 369.4s on 1.7M tables.
The linear increase in runtime for each corpus size is because the search space reduction percentage is stable on all synthetic corpora.
Runtimes using types are faster than using embeddings due to LSH using types filtering out a larger fraction of the search space: on average 91\% prefiltered using types and 74\% using embeddings.

\begin{figure}[t]
    \centering
    \includegraphics[width=0.8\columnwidth]{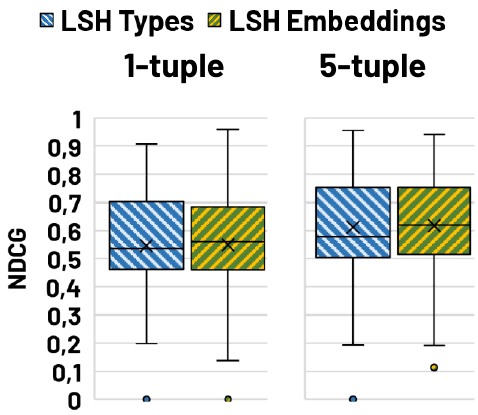}
    \caption{NDCG at top-100 for LSH prefiltering on WikiTables from 2019.}
    \label{fig:new_snapshot}
\end{figure}

\noindent
\textbf{Experiments on WikiTables from 2019.}
We also evaluate Thetis with LSH prefiltering using entity types and embeddings on the WikiTables from 2019, a dataset with more tables and a significantly lower entity link coverage.
The NDCG results are seen in Figure~\ref{fig:new_snapshot}.
These scores are very similar to those presented in
Figure~\ref{fig:ndcg}, showing that the accuracy of Thetis is not greatly influenced by a drop in coverage from 27.7\% to 18.2\% (see
Table~\ref{tab:benchmark}).
Due to the increased size of WikiTables from 2019, the runtimes on this corpus have increased to 26.2s and 35.1s on average using types and embeddings on 1-tuple queries, respectively, and 95.4s and 189.0s on 5-tuple queries.

\noindent
\textbf{Experiments on GitTables.}
We evaluate the runtime of \ourmethod on GitTables consisting of larger tables, i.e, more rows and columns (see Table~\ref{tab:benchmark}).
However, as this dataset does not come with ground truth, we do not evaluate the ranking quality on this dataset.
This dataset also does not come with entity link annotations.
Therefore, we construct Lucene indexes using the KG entity labels and perform keyword search to link mentions to KG entities.
The resulting runtimes are 1.7s and 3.7s on 1-tuple queries and 2.2s and 14.9s on 5-tuple queries.
These runtimes on GitTables are comparable to those corpora containing smaller tables, i.e., WikiTables from 2015 and 2019.
This is due to the LSH prefiltering reducing the GitTables corpus by more than 98\% for all queries, as the table entities are more evenly distributed across the LSH buckets.
Therefore, the LSH lookups are more selective on this dataset, meaning that \texttt{Thetis++} has to rank fewer tables than on WikiTables from 2015 and 2019.

\subsection{Varying Entity Linking Coverage}

\begin{figure}[t]
    \centering
    \includegraphics[width=\columnwidth]{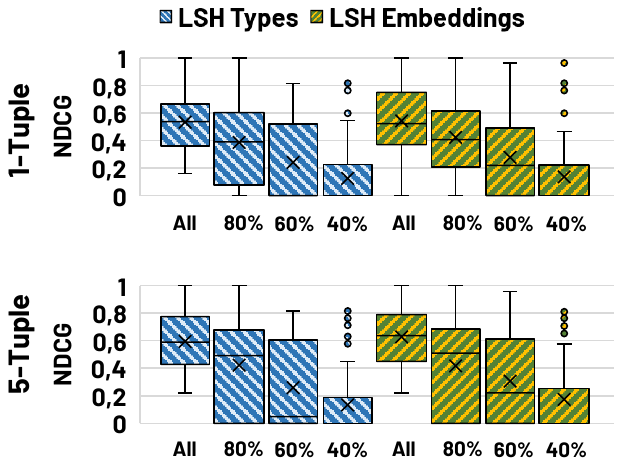}
    \caption{NDCG at top-10 when decreasing coverage}
    \label{fig:linking_coverage}
\end{figure}

Thetis exploits the contextual information available in the KG when entities in the tables are linked to it, while it does not require fully linked tables.
Here, we refer to \textit{link coverage} as the percentage of linked entities among all its cells.
Intuitively, the higher the coverage, the more information can be inferred about the contents of a table.
To maximize the coverage, it is important that the target KG can describe all the important entities in the data lake.
Nonetheless, it is very common to test open-domain KGs when trying to integrate data in data lakes~\cite{ilyas2023sagaextension}.
Our experiments have shown that our approach is competitive in precision to BM25 while being \emph{able to retrieve a large set of tables that were not retrieved otherwise}, even with less than 30\% of cells being linked to KG entities on average.
Therefore, even with partial overlap between the KG and the tables, Thetis is effective in retrieving relevant data.

We further experiment with different levels of table link coverage (Figure~\ref{fig:linking_coverage}). 
Specifically, we retrieve top-1000 tables and keep only tables with at most a given link coverage, e.g., only tables with link coverage up to 60\%.
We then evaluate NDCG on the top-10 of those retrieved tables.
As expected, the tables become increasingly more difficult to retrieve when entity link coverage decreases.
Yet, Thetis still retrieves relevant tables that BM25 cannot retrieve.
For 40\% link coverage, the median top-10 result set difference is 3 on 1-tuple queries using types and 2 using embeddings.
On 5-tuple queries, these numbers are 4 and 3, respectively.
There are still cases where, even when not many entities are linked, the method can achieve up to 0.8 of NDCG as it can capitalize on the semantic information of the few entities linked.

We also perform an experiment where we substitute the ground truth entity links in WikiTables from 2015 with the predicted entity links retrieved with a state-of-the-art entity linker (EMBLOOKUP~\cite{accelerating_linking}).
This leads to a mean linking coverage of only 20.3\% (compared to 27.7\% on WikiTables from 2015) and the F1-score of the employed entity linker is only 0.21.
Despite this, Thetis achieves an NDCG score of 0.14 using types and $k=10$ and 0.20 when using embeddings for 1-tuple queries and 0.26 and 0.29 for 5-tuple queries using types and embeddings, respectively.
This performance is better than the performance when decreasing the percentage of ground truth entity links to at most 40\% per table in Figure~\ref{fig:linking_coverage}.
Note that the 20.3\% is the mean coverage, whereas the 40\% in Figure~\ref{fig:linking_coverage} is an upper bound.
Therefore, Thetis is still able to retrieve meaningful results even when faced with poor entity linking quality.
\section{Conclusion} \label{sec:conclusion}
We have presented a scalable, semantic-aware table search algorithm, \texttt{Thetis++}.
\texttt{Thetis++} exploits a reference KG to facilitate entity-centric, exemplar querying for semantically related tables.
We have additionally studied and evaluated various techniques for combining rankings over data lakes for the data lake table search task.
We proposed a specific technique for combining rankings based on Pareto and TOPSIS and studied variations of the techniques involved in this approach.
We extended our study to alternative approaches using classification based on entity and entity type frequency, as well as a classification approach using entity embeddings.
Finally, we studied a learning-to-rank method and compared this to the evaluated methods.
We found that combining rankings is a difficult task, especially regarding the selection of true positives to be included in the combined top-\textit{K} rankings.
Therefore, the combining methods only obtain slightly improved ranking and output quality performance compared to the pure search engine performances.
However, learning-to-rank achieves stable recall in all evaluated cases, but outperforms the other evaluated approaches in ranking when increasing the query size.
As an alternative, Reciprocal Rank Fusion is left for future experimentation, as it has shown promising results in previous evaluations.

We proposed employing HNSW vector index for approximate nearest neighbor search and LSH to prefilter the search space.
HNSW prefiltering drastically improved the runtime performance of semantic table search in Thetis by 4.7$\times$ on average and up to 28$\times$, whereas LSH prefiltering achieves up to 4.2$\times$ better runtime performance.
It is even possible that one of the HNSW configurations improves the ranking performance of semantic table search in Thetis, as it removes false positives.
Our experiments show that complementing keyword search with \ourmethod allows one to find more relevant tables, improving recall by up to 5.4 times.
\section{Acknowledgements}
This research was partially funded by the Danish Council for Independent Research (DFF), grant agreement no. DFF-804800051B, the Poul Due Jensen Fond, and  DataGEMS, funded by the European Union's Horizon Europe Research and Innovation programme, grant agreement no. 101188416.
\section{Statements and Declarations}
There are no financial interests involved in this work.

\bibliographystyle{unsrt}
\bibliography{sample}

\end{document}